\documentclass[pre,preprint,showkeys,nofootinbib,preprintnumbers,amsmath,amssymb,floatfix]{revtex4-1}

\usepackage{graphicx}
\usepackage{dcolumn}
\usepackage{bm}
\usepackage{epsfig}
\usepackage{amssymb}
\usepackage{color}
\usepackage{footmisc}

\newcommand{\korr}{}
\newcommand{\tb}{}
\newcommand{\tr}{}
\newcommand{\tg}{}
\newcommand{\tmg}{}

\newcommand{\vect}[1]{\mathbf{#1}}
\newcommand{\kt}{k_{\rm B}T}

\newcommand{\jeansT}{{{\cal T}}}
\newcommand{\capL}{{\kappa}}
\newcommand{\be}{{\bf e}}

\newcommand{\br}{{\bf r}}
\newcommand{\bx}{{\bf x}}
\newcommand{\bk}{{\bf k}}

\newcommand{\bbg}{{\bf g}}
\newcommand{\jacob}{\mathcal{M}}

\begin{document}

\title{Capillary attraction induced collapse of colloidal monolayers at
    fluid interfaces}

\author{J. Bleibel$^{1,2}$, A. Dom\'\i nguez$^3$,
  M. Oettel$^2$, S. Dietrich$^{1,4}$} 
\affiliation{$^1$Max--Planck--Institut f\"ur Intelligente Systeme,
  Heisenbergstr.~3, 70569 Stuttgart, Germany} 
\affiliation{$^2$Institut f\"ur Angewandte Physik, Auf der Morgenstelle 10, Eberhard Karls
  Universit\"at, 72076 T\"ubingen, Germany}
\affiliation{$^3$F\'\i sica Te\'orica, Universidad de Sevilla, Apdo.~1065,
  41080 Sevilla, Spain}
\affiliation{$^4$IV.~Institut f\"ur Theoretische Physik,
  Universit\"at Stuttgart, Pfaffenwaldring 57, 70569 Stuttgart, Germany}

\date{\today}

\begin{abstract}
We investigate the evolution
of a system of colloidal particles, trapped at a fluid interface
and interacting via capillary attraction, 
as function of the range of the capillary interaction and
temperature. We address the collapse of an initially homogeneous
particle distribution and of a radially symmetric (disk--shaped) 
distribution of finite size, both theoretically by using a
perturbative approach inspired by cosmological models and
numerically by means of Brownian dynamics (BD) and dynamical density
functional theory (DDFT). The results are summarized in a ``dynamical
phase diagram'', describing a smooth
crossover from collective (gravitational--like) collapse
to local (spinodal--like) clustering.
In this crossover region, the evolution exhibits a peculiar 
shock wave behavior at the outer rim of the contracting, disk--shaped
distribution. 
\end{abstract}



\maketitle

\section{Introduction}
\label{sect:intro}

\tmg{Partially wetted colloidal particles can be strongly} trapped at fluid
interfaces. \tmg{There they form colloidal monolayers, which} constitute
systems of low \tmg{spatial dimensionality exhibiting a rich and interesting
  phenomenology. This is in part due to the effective capillary force acting
  between the particles, which is tied to the presence of a
  deformable fluid interface. \korr{Driven by basic research as well as
    application perspectives, capillary interactions have} recently received 
  renewed interest both in theory and in experiment, see,
  e.g., Refs.~\cite{Oettel:2008,Domi10,DaKr10a} and references
  therein. If the colloidal particles deform the interface as a
  consequence of an external force acting on them, such as their weight
  or an externally imposed electric field among several actual experimental
  realizations, the dominant contribution to the 
  capillary interaction is the so-called ``capillary
  monopole''{\renewcommand{\baselinestretch}{1.0}\normalsize\footnote{In the absence of an external force, the
    dominant contribution is the ``capillary quadrupole'', which is
    related to deviations from spherical symmetry of the particles. See, e.g.,
    Ref.~\cite{BLCS12} for a recent review.}}. In the (experimentally
  realistic) limit of small interfacial deformations, this force is
  always attractive and has a range given by the}
capillary length $\lambda=\sqrt{\gamma/(g \Delta\rho)}$, where
$\gamma$ is the surface tension of the interface, $g$ is the
acceleration of gravity, and $\Delta\rho$ is the mass density
difference \tmg{between the fluids on both sides of the interface. In
  typical experimental configurations the interparticle separation
  ($\sim$ micron) is much \tr{smaller} than $\lambda$ ($\sim$
  millimeter), so that any  
  \tg{given} particle interacts simultaneously with
  a huge number of neighbors. Furthermore, at separations smaller
  than $\lambda$, the capillary attraction varies logarithmically and
  is formally analogous to two--dimensional (2D) Newtonian gravity.
  In these circumstances, the capillary interaction is \korr{equivalent to}
  ``screened 2D 
  Newtonian gravity'', with a screening length $\lambda$ much larger
  than any microscopic length scale. Therefore we call this interaction
  ``long--ranged''}. In this context the \emph{tunability} of such a system is
of special interest. The range of the interaction
may in principle be varied 
by changing the capillary length, which is accessible via changes of
the surface tension of the interface, e.g., due to the addition of
surfactants. \tmg{Additionally, the relevant ratios of the interaction range to
  other length scales of the system (most notably, the interparticle
  separation and the system size) can be easily altered also by
  changing these other lengths in experimental setups. Likewise, the
  amplitude of the capillary monopole and, correspondingly, the
  strength of the interaction is controllable by means of the external
  force. Therefore, } for micron--sized colloidal
particles, \tmg{such a monolayer} can serve as a 2D model system for
\tmg{studying the \korr{influence} of a ``long--ranged'' interaction as a
  function of its parametric characteristics}.

Long--ranged interactions are a subject of active research in various
branches of physics (see, e.g., Ref.~\cite{CDR09} and references
therein), involving rather distinct scales. Ranging from cosmology
down to chemotaxis in bacterial populations, a common feature of these
systems is their temperature dependent instability with respect to
\tmg{clustering}
\cite{Keller:1970,ChSi08,Chavanis:2010,Dominguez:2010}.
In recent studies, we investigated \tmg{this clustering instability in monolayers} of micron sized 
particles \tmg{driven by \korr{a} monopolar} 
capillary
interaction~\cite{Dominguez:2010,Bleibel:L2011,Bleibel:2011},
\tmg{paying special attention to the influence of the ratio between the
  capillary length and the system size. In the formal limit
  $\lambda\to\infty$}, corresponding to 2D Newtonian gravity,
cosmology offers a \tmg{useful analogy \korr{for tackling} the problem within}
the so--called ``cold collapse'' approximation, which is widely used to study
clustering of 
collisionless matter: \tmg{one neglects any force other than gravity
  (including diffusion by thermal motion). Within this approximation, the
  dynamical equations of our model are amenable to an analytical
  solution, which provides the basis for perturbative calculations
  when $\lambda$ is finite. Numerical solutions and N--body simulations
  allow for going beyond the simplifying assumptions
  of the analytical approach.}

\tmg{Here our goal is to provide a complete characterization of the
  evolution induced by the clustering instability as a function of the
  relevant parameters of the system. In this respect in
  Refs.~\cite{Dominguez:2010,Bleibel:2011} a homogeneous particle
  distribution (either macroscopically extended or with periodic boundary
  conditions) was studied, while Ref.~\cite{Bleibel:L2011} \tr{has} addressed a
  radially symmetric (disk--shaped) distribution with a finite radius, which
  is more 
  realistic from an experimental point of view. There a 
  \tr{``dynamical phase diagram''} was proposed in order to summarize the findings
  (see Fig. 1 in Ref.~\cite{Bleibel:L2011}): there is a region of
  stability at high temperatures, and a region of instability at low
  temperatures. In the latter region, the evolution proceeds as in the case of
  gravitational cold collapse if $\lambda$ is larger than the system
  size, and as in the case of spinodal decomposition if $\lambda$ is much
  smaller 
  than the system size. There is no sharp border line between these two
  limiting cases but instead a crossover region which exhibits characteristic
  features of both regimes. In a \tr{disk--shaped} particle distribution,
  the presence of the explicit outer boundary brings into focus this crossover
  region as in the course of the collapse a shockwave--like structure is
  formed. First corresponding results were communicated in
  Ref.~\cite{Bleibel:L2011}. Here we
  elaborate and extend the theoretical model and the perturbative
  calculations underlying this study\tr{,} including the proposed ``dynamical phase
  diagram''.}
Resorting to Brownian dynamics (BD) simulations and
dynamic density functional theory (DDFT) we investigate this diagram
concerning the dynamic evolution of a \tmg{disk--shaped} finite distribution
of particles.
\tmg{Both the BD and the DDFT results support the picture captured by the
  \tr{``dynamical phase diagram''}.}

The paper is organized such that in the next section we shall \tmg{present
  the theoretical model. In Subsec.~\ref{sect:theory} we shall 
 introduce the definitions and the terminology used throughout the paper}. We
\tmg{first} recall briefly the 
theory for colloidal systems with \tmg{monopolar} capillary
attractions \tmg{and the model for the dynamical evolution}.
\tmg{Next, in Subsec.~\ref{sec:hom} we shall discuss the} linear stability
analysis of the homogeneous system, \tmg{based on which we identify
  various dynamical regimes} in the \tr{``dynamical phase diagram''} \tb{as a
  function of the 
  screening length $\lambda$ and a rescaled, effective temperature}. This will
be followed by a formal definition of the cold collapse model, which
serves as the baseline for the formulation of an analytical
perturbation theory for the capillary collapse of colloids trapped at
a fluid interface. \tb{The main results will be summarized in
  Subsec.~\ref{sec:pert_summary} and certain derivations are given in
  App.~\ref{app:pert}.} \tmg{In Sec.~\ref{sect:2DDDFT} we shall discuss} the
numerical methods \tmg{we have applied in order to investigate the
  capillary collapse}. 
In Sec.~\ref{sect:result} we shall report on simulations for various setups so
as to explore the ``dynamical 
phase diagram'' also for the case of finite--sized circular patches of
particles, as function of an effective temperature and of the range of the
interaction. In order to render a
comparison with \korr{future} experiments feasible, \korr{in
Sec.~\ref{sect:lambda}} we shall briefly discuss  how a change
of the interface tension affects the relevant variables
for the setup of a homogeneous system or for that of a collapsing
disk. \tmg{Finally}, \tr{in Sec.~\ref{sect:sum}} we shall summarize and
present our  conclusions.

\begin{figure}
\centerline{\epsfig{file=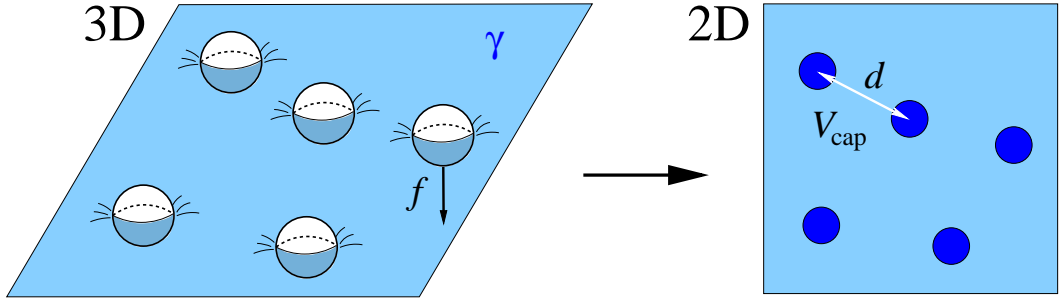, width=14cm}}
\caption{\tb{Sketch of the system under consideration. Colloids are trapped at
    the interface between two fluids with surface tension $\gamma$.
Around each colloid a dimple is formed because the external force $f$ acting
on each colloid pushes the colloids into the lower fluid (often water). The
dimple depth is proportional to $f/\gamma$. Thus for micron--sized colloids
with $f$ caused by their own weight (or buoyancy) the depth is in the range of
a few nm. Therefore the system is effectively two--dimensional,
and the colloids interact via the attractive, long--ranged capillary pair
potential $V_\mathrm{cap}(d)$ (\tmg{see Eq.~(\ref{eq:2part})}) in addition to
short--ranged, repulsive interactions of different nature. This pair potential
constitutes the dominant term of the colloidal interaction within \tmg{the
  multipolar expansion} which amounts to an expansion  
in terms of \tmg{the inverse interparticle separation} $1/d$
\cite{ODD05,Dominguez:2008,Domi10}.} The actual three--dimensional \tr{(3D)}
  configuration is mapped onto a two--dimensional \tr{(2D)} configuration of
  disks in a reference plane.} 
\label{fig:system}
\end{figure}

\section{Theoretical model}
\subsection{Basic features and description of the dynamic evolution}
\label{sect:theory}

We consider a two--dimensional set of interacting particles   
(Fig.~\ref{fig:system}). For the system under study, we map a distribution of
spherical colloidal particles \tr{of radius $R$,} which are trapped at a
fluid interface\tr{,} onto a two--dimensional distribution of disks in a flat
reference plane. For  
small deformations of the interface as considered here, this renders a
two--dimensional distribution of circular disks with radius
$R_0$~\cite{ODD05,Dominguez:2008,Oettel:2008}. \tmg{(The relationship
  between $R$ and $R_0$ depends on the contact angle of the
  interface at the particle surface.)} The dynamical model \korr{of} them as
used 
here \tr{has been} introduced in Ref.~\cite{Dominguez:2010} and is
briefly recalled here.

The particles are subject to  
capillary forces. These are
generated by the interfacial deformation \tb{upon the action of an
external force $f$ on the particles (which in its turn is, e.g., caused by
the weight of the particles and their buoyancy at the interface) 
in the direction
perpendicular to the reference plane}. Within mean--field theory, the \tb{linearized} Young--Laplace equation governing the interfacial
deformation $U(\hat{\br})$, $\hat{\br}\in\mathbb{R}^2$, is given
by~\cite{ODD05,Oettel:2008}
\begin{equation}
  \label{eq:YL}
  \hat{\nabla}^2 U - \frac{U}{\lambda^2} = - \frac{f}{\gamma} \varrho ,
\end{equation}
where $\varrho(\hat{\br})$ is the 2D particle number density field in the
reference plane and $\gamma$ is the surface tension. \tb{(The interfacial
  deformation for micron--sized particles
is in the nm range, which justifies the linearized approximation.)}
This interfacial deformation gives rise to an in--plane capillary
force described by the areal force density 
\begin{equation}
  \label{eq:fcap}
  \vect{F}_{\rm cap}=f\varrho \hat{\nabla} U .
\end{equation}
\tr{The capillary force between two particles can be cast into a form such that
a corresponding pair potential is given by~\cite{Oettel:2008}}
\begin{equation}
  \label{eq:2part}
  V_\mathrm{cap}(d) = - \frac{f^2}{2\pi\gamma}  
  K_0 \left(\frac{d}{\lambda}\right) 
\end{equation}
in terms of the modified Bessel function $K_0$, as obtained from
  Green's function for Eq.~(\ref{eq:YL})\tr{, and the
    \korr{center--to--center} distance $d$ between the
  particles}. The capillary length
$\lambda$ acts as a cutoff for the interaction range because
$K_0(d/\lambda\to \infty) = \bigl(\frac{2}{\pi} \frac{d}{\lambda}\bigr)^{-1/2}
\mathrm{e}^{-d/\lambda}$. In the opposite limit, $d\ll \lambda$, this potential
coincides with the Newtonian gravitational interaction in $d=2$,
$K_0(d/\lambda\to 0) = \ln (\lambda/d)$, \tmg{which is nonintegrable in
  the sense of statistical mechanics, i.e., it leads to a nonadditive
  (hyperextensive) internal energy. In view of this property and due to the
  wide separation of scales in colloidal systems, $\lambda =
  \mathcal{O}(\mathrm{mm})$ and $R_0=\mathcal{O}(\mu\mathrm{m})$,
  for the present purposes we call  $V_\mathrm{cap}(d)$ a ``long--ranged
  interaction'', as opposed to other interactions which will be called
  ``short--ranged''}. 

Concerning the dynamics, we assume an overdamped motion for
the particles adsorbed at the interface. For reasons of
simplicity, \tr{as a first step} we neglect hydrodynamic
interactions{\renewcommand{\baselinestretch}{1.0}\normalsize\footnote{This is justified 
  for dilute systems \tr{(see,
    c.f., Refs.~\cite{Dominguez:2010,Bleibel:2011}). For} a 
  similar system with hydrodynamic interactions included see
  Refs.~\cite{Bleibel:L2011,Bleibel:L2013}.}}, so that the in-plane number
density current of the particles is proportional to the driving force,
\begin{equation}
  \label{eq:drag}
  \varrho\vect{v} 
  =\Gamma\, 
  (\vect{F}_{\rm short}+\vect{F}_{\rm cap} )
\end{equation}
with an effective single--particle mobility $\Gamma$. 
$\vect{F}_{\rm short}$ is the sum of the areal \tmg{thermodynamic force
  density associated with 
  thermal motion and} of any other (short--ranged) interaction between
the particles. Assuming local equilibrium, one can write
$\vect{F}_{\rm short} = - \hat{\nabla} p$, where the 2D pressure field
$p=p(\varrho)$ is given by an appropriate equilibrium equation of
state describing the macroscopic manifestation of the short--ranged forces.
Therefore, the continuity equation for the particle number density is
\begin{equation}
  \label{eq:cont}
  \frac{\partial \varrho}{\partial \hat{t}}=-\hat{\nabla}\cdot\left(\varrho \vect{v}\right)=
  \Gamma\hat{\nabla}\cdot\left(\hat{\nabla} p-f\varrho \hat{\nabla} U\right) .
\end{equation}
We introduce a characteristic particle density $\varrho_0$ and \tb{the length
  scale $L_0$ associated with the size of the system. (For the homogeneous
  systems considered below, $L_0$ is the
side length of the periodically replicated square box and for the collapsing
disk scenario, it 
is the initial radius of the particle--covered disk.)
} 
We also introduce the time scale $\jeansT := \gamma /( \Gamma f^2 \varrho_0)$,
which is the so-called Jeans' time associated with the \tr{initial} homogeneous
density $\varrho_0$ \cite{Dominguez:2010}. This allows us to form the
following dimensionless variables \korr{and quantities}:
\begin{subequations}
  \begin{equation}
    \br := \frac{\hat{\br}}{L_0} , \quad 
    t := \frac{\hat{t}}{\jeansT} , \quad 
    \capL := \frac{L_0}{\lambda} , \quad
    n(\br, t) := \frac{\varrho(\hat{\br}, \hat{t})}{\varrho_0} , 
  \end{equation}
  \begin{equation}
    \label{eq:dimless}
    w(\br, t) := \frac{\gamma}{f \varrho_0 L_0^2} U(\hat{\br}, \hat{t}) , \qquad
    \Pi (n) := \frac{\gamma}{f^2 \varrho_0^2 L_0^2} p(\varrho) .
  \end{equation}
\end{subequations}
In these terms, Eqs.~(\ref{eq:YL}) and (\ref{eq:cont}) take the form
\begin{subequations}
  \label{eq:euler}
\begin{equation}
  \label{eq:w}
  \tr{\nabla^2 w - \capL^2 w = - n \,,}
\end{equation}
\begin{equation}
  \label{eq:n}
    \frac{\partial n}{\partial t} = -\nabla
    \cdot[ n \nabla w - \nabla \Pi ] .
\end{equation}
\end{subequations}
\tmg{It is useful to introduce the (dimensionless) chemical potential
  $\mu=\hat{\mu}\gamma/(f^2\varrho_0 L_0^2)$ which is 
  associated with the pressure via the Gibbs--Duhem
  relation{\renewcommand{\baselinestretch}{1.0}\normalsize\footnote{The additive function of temperature only 
    is irrelevant for our present purposes, given that the system evolves under
    isothermal conditions.}}
  \begin{equation}
    \label{eq:gibbsduhem}
    \frac{\partial \mu}{\partial \Pi} = \frac{1}{n}
    \quad\Rightarrow\quad
    \mu(n,T) = \int_{n^{*}}^n \frac{dn^{'}}{n^{'}} \frac{d\Pi}{d n^{'}} +
    \mu(n^{*},T) ,
  \end{equation}
  \tr{with a certain constant $n^{*}$} so that Eq.~(\ref{eq:n}) can be also written
  as 
\begin{equation}
  \label{eq:nwithmu}
    \frac{\partial n}{\partial t} = \nabla
    \cdot[ n \nabla (\mu-w) ] .
\end{equation}
Alternatively, one can introduce a free energy functional
$\mathcal{F}[n,w]$ (see Ref.~\cite{Dominguez:2010}), such that
\begin{equation}
  \frac{\delta \mathcal{F}}{\delta n} = \mu - w ,
\end{equation}
so that Eqs.~(\ref{eq:w}) and (\ref{eq:n}) can be written as
\begin{equation}
  \label{eq:ddft}
  \frac{\delta \mathcal{F}}{\delta w} = 0 ,
  \quad
  \frac{\partial n}{\partial t} = \nabla
  \cdot\left[ n \nabla \frac{\delta \mathcal{F}}{\delta n} \right],
\end{equation}
respectively.
This establishes a formal analogy between Eq.~(\ref{eq:euler}) and
the so called \textit{dynamic density functional theory} (DDFT)
\cite{Mar99} for the ensemble--averaged density.}

The dynamics described by these equations has been already the subject
of previous studies
\cite{ChSi08,Dominguez:2010,Bleibel:L2011,Bleibel:2011,Chav11}. The
main goal of the following analysis is the investigation of the dynamics
as a function of the capillary length 
\tr{if} the initial condition is a circular patch of colloidal particles,
i.e., the solution of these equations with \korr{the property that $w(\br, t)$
  is regular and with} the boundary condition
\begin{equation}
  \label{eq:bc}
  \lim_{r\to\infty} w(\br, t) = 0 ,
\end{equation}
and a so--called ``top--hat''profile as initial condition\tr{:}
\begin{equation}
  \label{eq:n0}
  n_0 (\br) := n(\br, t=0) = \left\{
    \begin{array}[c]{cl}
      1 , & r < 1 \\
      & \\
      0 , & 1 < r . \\
    \end{array} \right. 
\end{equation}
\tmg{This initial configuration complements and is
  closer to experimental realizations } 
than the idealized case of an infinitely extended homogeneous
distribution. Compared with the cases of an infinitely extended system or
a system in a box with periodic boundary conditions (as addressed in
Refs.~\cite{ChSi08,Dominguez:2010,Bleibel:2011,Chav11}), the presence
of an explicit boundary brings into focus the effect of \tmg{changing
  the range of the capillary attraction relative to the system size,} 
leading to a peculiar phenomenology~\cite{Bleibel:L2011}.

For the following discussion we introduce the useful notions of the ``cold
limit'' ($\Pi \to 0$ in Eq.~(\ref{eq:n}) \tmg{or $\mu\to 0$ in
  Eq.~(\ref{eq:nwithmu})}) and of the ``Newtonian limit'' ($\capL \to 0$ in
Eq.~(\ref{eq:w})). The first notion refers to the interpretation of
vanishing pressure as a sort of athermal limit of the fluid; the second
notion refers to the case that the capillary attraction
\tmg{(Eq.~(\ref{eq:2part}))} becomes identical to Newtonian
gravitational attraction in 2D \tmg{(see, e.g.\tr{,}
 Refs.~\cite{Dominguez:2010,ChSi11})}. Regarding this latter notion, \tmg{we
  briefly remark on the opposite limit $\capL\to\infty$, i.e., a very
  short--ranged capillary attraction. From the explicit integral of
  Eq.~(\ref{eq:w}),
\begin{equation}
  w(\br) = \frac{1}{2\pi} \int_{\mathbb{R}^2} d^2\br'\; n(\br') K_0 (\capL |\br-\br'|) ,
\end{equation}
one deduces 
\begin{equation}
  \label{eq:largekappa}
  w \approx \frac{n}{\capL^2}, 
  \qquad
  \capL \gg 1,
\end{equation}
in regions where the density field is smooth,}
and Eq.~(\ref{eq:n}) becomes 
\begin{equation} 
  \label{eq:diff_large_kappa}
  \frac{\partial n}{\partial t} \approx \nabla^2 P ,
  \qquad P(n) = \Pi(n) - \frac{n^2}{2 \capL^{2}} ,
\end{equation}
\tb{with $\nabla(n\nabla w) \approx \kappa^{-2} \nabla(n\nabla n) =
  (\kappa^{-2}/2) \nabla^2(n^2)$ }.
This is the diffusion equation for a fluid the equation of state
$P(n)$ of which accounts for the attractive force through the van der
Waals--like contribution $-n^2/(2\kappa^2)$.
\tmg{In 
  \tg{view of that}, one should note that, motivated by the wide scale
  separation mentioned above ($\lambda\gg R_0$) we are interested in
  the comparison of the range of the interaction with a
  characteristic system size, as quantified by the parameter
  $\capL$. This is  
  \tg{distinct from} the more usual approach, in which the range
  would be compared with a microscopic length, e.g., the particle
  radius.}

\subsection{Stability with periodic boundary conditions}
\label{sec:hom}

We first review briefly the linear stability analysis of a homogeneous
distribution in a finite \tb{quadratic} box with \tb{side length $L_0$
  and} periodic boundary conditions. \tmg{The findings can be
  summarized in a \tr{``dynamical phase diagram''}, which is also useful} 
for understanding qualitatively the mathematically more
difficult case of a finite disk.

\begin{figure*}
  \hfill\epsfig{file=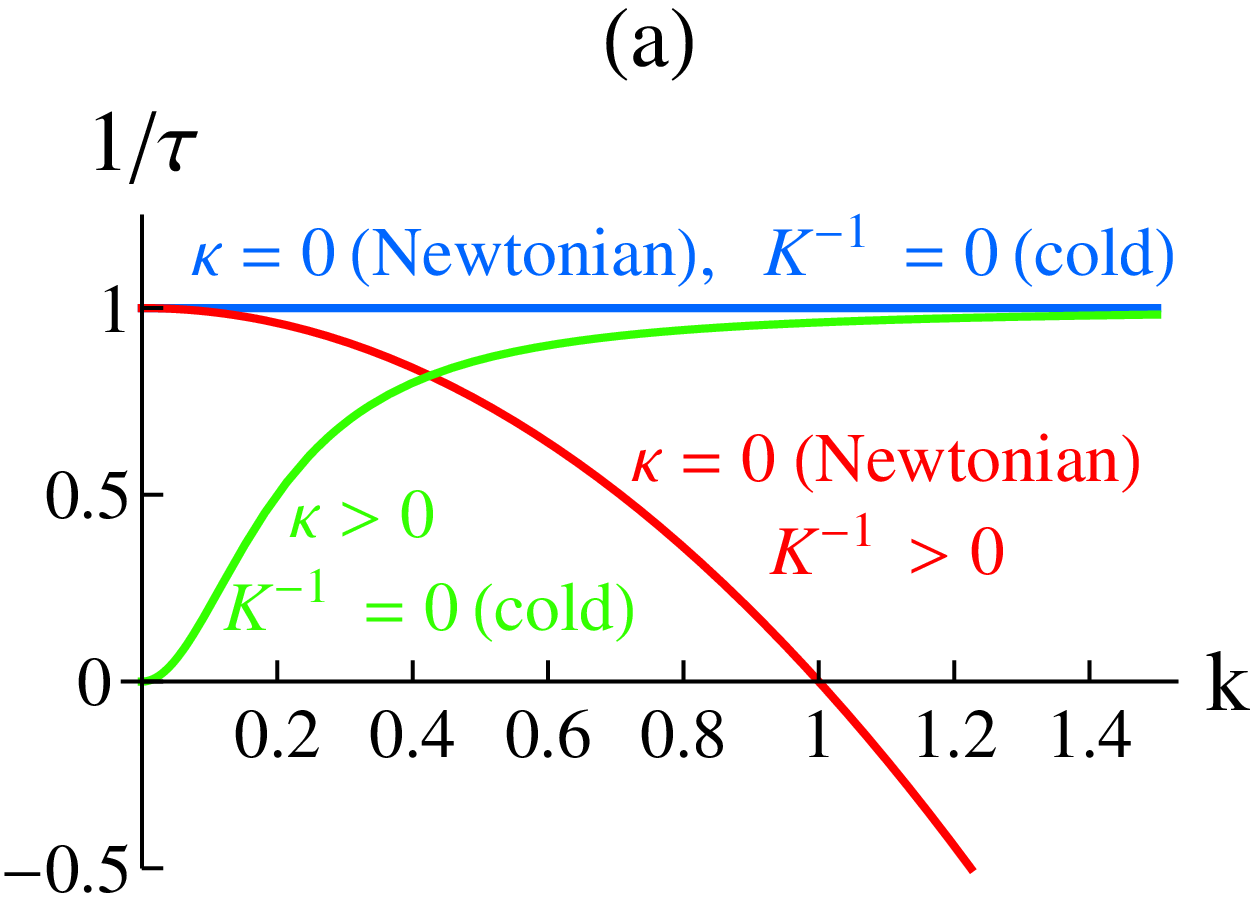,width=.4\textwidth}
  \hfill\epsfig{file=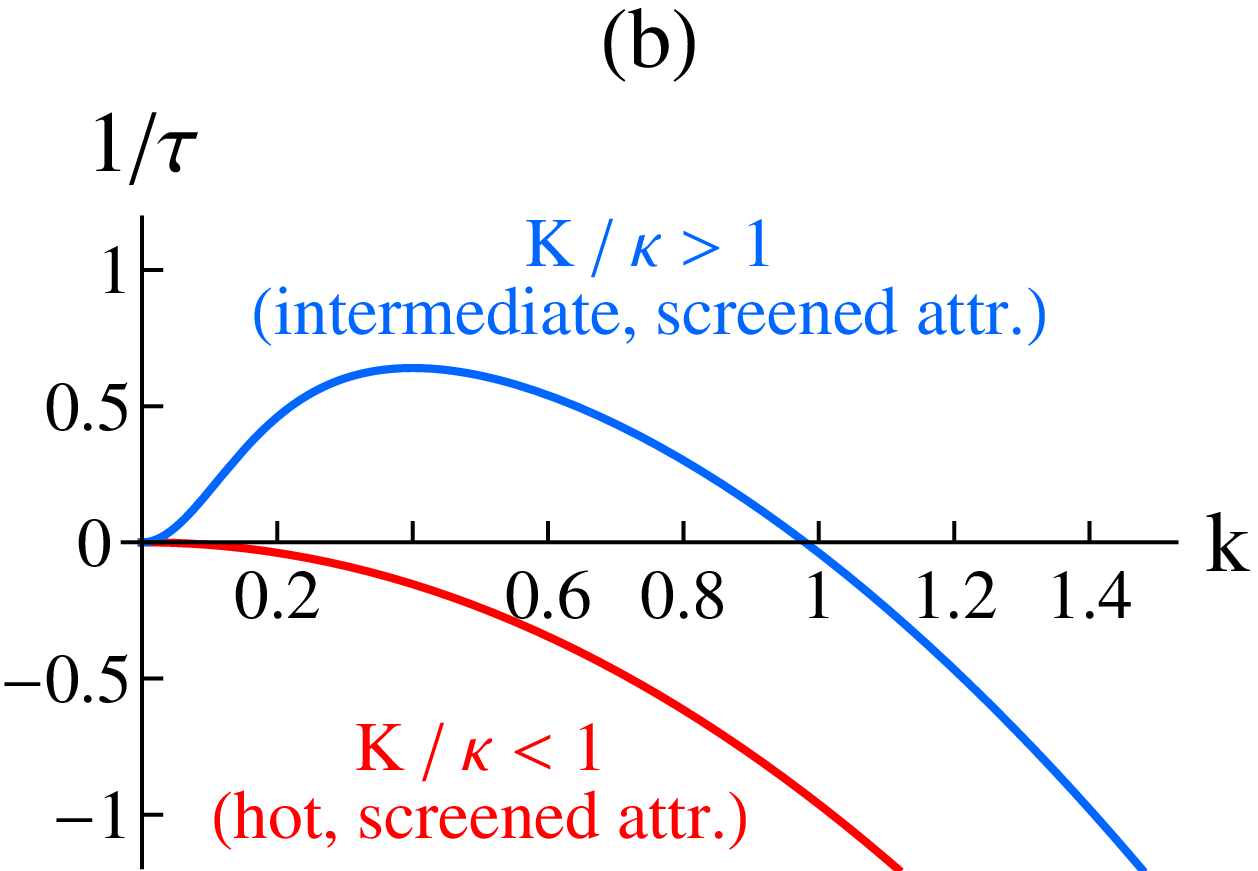,width=.4\textwidth}
  \hspace*{\fill}
  \caption{\label{fig:tau} \tr{Examples of the dependence on
    $k$ of the growth rate $1/\tau(k)$ defined in Eq.~(\ref{eq:tau})
    for various values of the parameters $K$ and $\capL$ (see the main
    text). In (a), $K^{-1}=1$ for the red line and $\capL=0.2$
    for the green line; in (b), $K^{-1}=1$ \korr{and} $\capL=5$ for the red
    line, and $K^{-1}=1$ \korr{and} $\capL=0.2$ for the blue line.}}
\end{figure*}

Consider the equilibrium, homogeneous state $n_\mathrm{eq}(\br) = 1$
with $\br\in [0,1]\times[0,1]$ and periodic boundary conditions.  
{In order to account for a locally perturbed density 
  $n(\br, t) = 1 + \delta(\br, t)$ one introduces the Fourier transform
of the} density deviations 
\begin{equation}
  \tilde{\delta}_\bk (t) := \int_{[0,1]^2} d^2\br\; \delta(\br,t) \mathrm{e}^{-i\bk\cdot\br}
\end{equation}
with the wavevectors (in units of $1/L_0$)
\begin{equation}
  \bk = 2\pi {(m_x, m_y)}, 
  \qquad
  m_x,m_y \in\mathbb{Z}.
\end{equation}
Equation~(\ref{eq:euler}) can be linearized in $\delta$ resulting in
\begin{equation}
  \label{eq:deltalin}
  \tilde{\delta}_\bk (t) = \tilde{\delta}_\bk (0) \mathrm{e}^{t/\tau(k)} ,
\end{equation}
where in units of $\mathcal{T}$ the \tb{inverse} characteristic time {for each
  mode $\bk$ } is given {by} 
\begin{equation}
  \label{eq:tau}
  \frac{1}{\tau(k)} = k^2  
  \left[
    \frac{1}{k^2 + \capL^2} - \frac{1}{K^2}
  \right] ,
\end{equation}
in terms of Jeans' length {$K^{-1}$} (in units of $L_0$) defined {as}
\begin{equation}
  {K^{-2} := \left.\frac{d \Pi}{d n}\right|_{n=1} \equiv \Pi^{'}(n=1).} 
\end{equation}
\tmg{(This quantity is actually the inverse of the isothermal
  compressibility of the fluid due to thermal motion and
  short--ranged forces, i.e., in the absence of capillary
  attraction.)}  Figure~\ref{fig:tau} {depicts} $1/\tau(k)$ for two
qualitatively distinct cases characterized by the values of $K$ and
$\capL$. In the cold ($K\to\infty$) Newtonian ($\capL=0$) limit all
modes grow at the same rate, i.e., \tb{$\tau(k)\equiv 1$} (horizontal blue
line in Fig~\ref{fig:tau}(a)).
\tmg{(Actually, in this limit the growth of perturbations can be
  computed exactly beyond the linear regime; see the discussion in
  \tr{Subsec.}~\ref{sec:pert_summary}).} \tb{A nonzero} pressure ($K<\infty$)
causes a slowing down \tb{of the growth} of the small--scale
($k\to\infty$) perturbations or even leads to damping 
\tb{($1/\tau(k)<0$)} \tmg{due to the finite compressibility} (red line in
Fig~\ref{fig:tau}(a)). The 
effect of a finite range of {the} attraction ($\capL>0$) is to slow
down
the growth of the large--scale ($k\to 0$) perturbations (green line in
Fig~\ref{fig:tau}(a)). If $K/\capL <
1$, all perturbations are damped (red curve in
Fig~\ref{fig:tau}(b)); if $K/\capL > 1$, the amplitude of all
Fourier modes \tb{below} a critical wavenumber grow in time (blue curve in
Fig~\ref{fig:tau}(b)).

\tmg{The analysis is facilitated by the introduction of a
  dimensionless parameter which we call ``effective temperature'':}
\begin{equation}
  \label{eq:Teff}
  T_\mathrm{eff} := \frac{(2\pi)^2+\capL^2}{K^2} .
\end{equation}
This parameter quantifies the 
ratio \tmg{between the energy associated with thermal motion and short--ranged
  forces and \tr{with}}  
the potential energy of the capillary attraction.

Indeed, if for simplicity one considers an ideal gas, $p(\varrho) =
\varrho \tb{\kt}$, in physical units Eq.~(\ref{eq:Teff}) turns into
\begin{equation}
  \label{eq:Teffideal}
  T_\mathrm{eff} = \frac{\gamma \tb{\kt}}{f^2 N_\mathrm{neigh}} ,
\end{equation}
where the capillary potential energy of a pair of particles is $\sim
f^2/\gamma$ (see Eq.~(\ref{eq:2part})) and 
\begin{equation}
  N_\mathrm{neigh} = \varrho_0 \left[ 
    \left(\frac{2\pi}{L_0}\right)^2 + \frac{1}{\lambda^2}
  \right]^{-1} 
\end{equation}
\tmg{gives the approximate} average number of neighbors with which
any  
\tg{given} particle interacts via capillary attraction. \tmg{This quantity
  allows one \tg{to} quantify the notions of a ``large system'', for which
  $N_\mathrm{neigh} \sim \varrho_0 \lambda^2$ if $\capL = L_0 /
  \lambda \gg 1$, and of a ``small system'', with $N_\mathrm{neigh} \sim
  \varrho_0 L_0^2$ if $\capL = L_0 / \lambda \ll 1$.}

\begin{figure*}
  \psfig{file=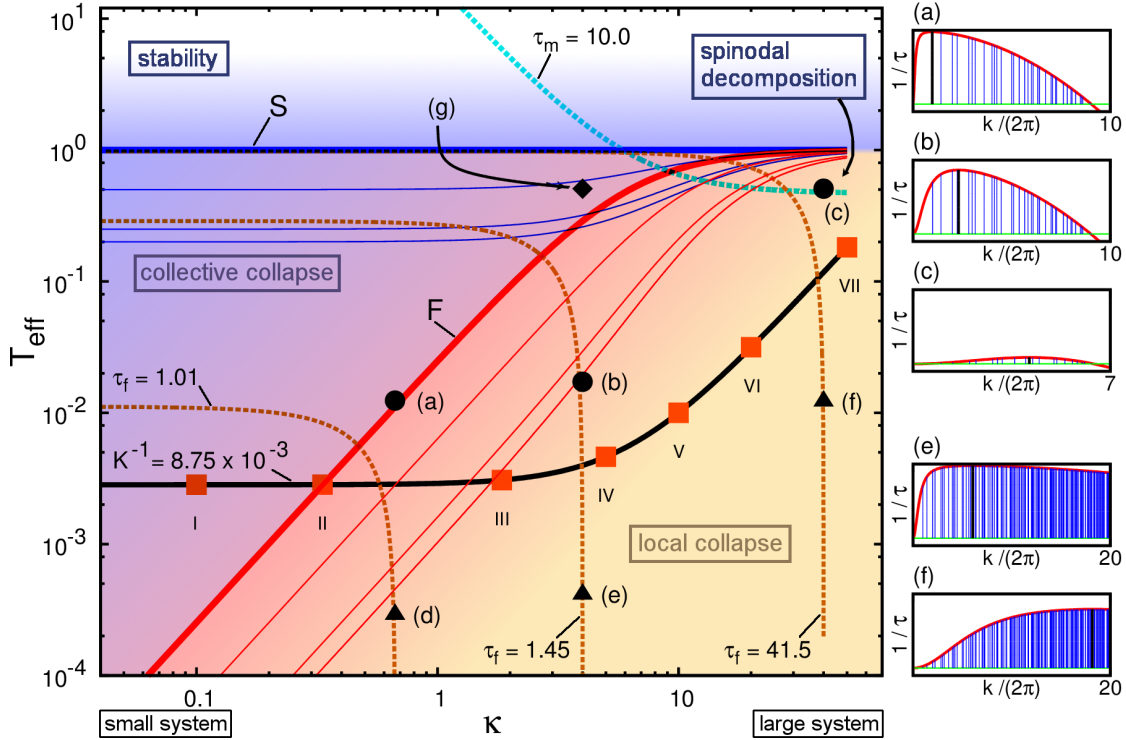,width=0.9\linewidth}
  \renewcommand{\baselinestretch}{1.4}\normalsize
  \caption{\label{fig:phase} Diagram showing the different ``regimes
    of instability'' for an initially homogeneous distribution with periodic
    boundary conditions in terms of $T_\mathrm{eff}$ (Eq.~(\ref{eq:Teff}))
    and $\capL=L_0/\lambda$. {\em Blue lines}, depending on the
    squared modulus $m^2=m_x^2+m_y^2$ of the mode indices, indicate 
    the limit of stability for those modes (Eq.~(\ref{eq:kunstable})).
    The thick blue line S of absolute stability
    corresponds to $m^2=1$, subsequent lines correspond to
    $m^2=2$, $m^2=4$, and $m^2= 5$. {\em Red lines} indicate
    that a mode with squared modulus $m^2$ is the fastest growing
    one (Eq.~(\ref{eq:kmaxgrowth})). The thick red line F corresponds to
    $m^2=1$. The thick {\em black line} is an iso--$K$ line
    \tr{$T_{\rm{eff}}(\capL)=K^{-2} \bigl [(2\pi)^2+\capL^2 \bigr ]$}
    (Eq.~(\ref{eq:Teff})), containing the 
    state points of the 
    Brownian Dynamics simulations presented in
    Ref.~\cite{Bleibel:2011} with a Jeans' length of $K^{-1} = 8.75
    \times 10^{-3}$ (filled squares).  Finally, the lines
    $\tau_\mathrm{m}=10$ (i.e., the fastest mode has \tr{the} characteristic
    time 10 
    $\jeansT$) and $\tau_\mathrm{f}=\tg{(1.01,\,1.45,\,41.5)}$ (from
    left to right; the fundamental mode has the characteristic time
    $1.01\jeansT,\, 1.45\jeansT,\,\rm{and }\,41.5 \jeansT$, respectively)
    are given by  
    Eq.~(\ref{eq:taum}) and Eq.~(\ref{eq:tauf}), respectively.
    The lines for given values of $\tau_\mathrm{f} \gtrsim 1$
    indicate the transition region. The black symbols (symbols
    \tr{of different shape} correspond to different Jeans' lengths) 
    indicate the systems consisting of a finite circular initial 
    patch of particles investigated via simulations and discussed
    in Sec.~\ref{sect:result}. For the points (a) - (c) and (e),
    (f),  the corresponding inverse characteristic times $1/\tau(k)$ (as in
    Fig.~\ref{fig:tau}, see Eq.~(\ref{eq:tau})) with the discrete modes
    indicated by blue vertical lines are shown in the side panels on the
    right. Note that the horizontal scales for $k$ vary whereas the
    vertical scale always ranges from $-0.1 < 1/\tau < 1$. The green
    horizontal line corresponds to $1/\tau = 0$ and the \tr{vertical} black
    lines indicate the maxima of $1/\tau(k)$.}
\end{figure*}

In terms of these
parameters $T_{\mathrm{eff}}$ (Eq.~(\ref{eq:Teff})) and $\capL$, for a general system
Eq.~(\ref{eq:tau}) {becomes}
\begin{equation}
  \frac{1}{\tau(k)} \tb{\equiv  \frac{1}{\tau(k; \kappa, T_\mathrm{eff})}} = \frac{k^2}{(2\pi)^2 + \capL^2} \left[
    \frac{(2\pi)^2 + \capL^2}{k^2 + \capL^2} - T_\mathrm{eff}
  \right] ,
\end{equation}
so that the homogenous state is linearly stable ($\tau(k)<0$ for all allowed
values of $k$) if\tmg{ and only if} $T_\mathrm{eff}>1$. As
$T_\mathrm{eff}$ decreases, more and more modes at smaller length scales
become unstable. This is visualized in 
Fig.~\ref{fig:phase}, which facilitates the discussion of the
evolution of the density field in real space as function of the
two parameters $T_\mathrm{eff}$ and $\capL$ characterizing the initial
state. 
We have found that two properties are relevant {in order to identify}
different ``regimes of instability'': first, \tr{the number of unstable} Fourier
modes, as determined by the position of the zero in
Fig.~\ref{fig:tau}, and second, \tr{the kind of} {mode \tr{exhibiting} the
  fastest 
  growth,} determined by the position of the maximum in Fig.~\ref{fig:tau}.
\begin{enumerate}
\item The limit of stability in Fig.~\ref{fig:phase} of a given mode
  $\bk = 2\pi {(m_x,m_y)}$ is determined by the condition
  \begin{equation}
    \label{eq:kunstable}
    \frac{1}{\tb{\tau(k;\kappa,  T_\mathrm{eff})}} = 0
    \quad\Rightarrow\quad
    \tb{T_\mathrm{eff}(m_x,m_y;\kappa)} = \frac{1 + (\capL/(2\pi))^2}{m_x^2+m_y^2 +
      (\capL/(2\pi))^2} . 
  \end{equation}
  In particular, for the fundamental mode ($k=2\pi$) this defines the
  line $T_\mathrm{eff}=1$ of absolute stability (``line S'' in
  Fig.~\ref{fig:phase}).
\item The loci in Fig.~\ref{fig:phase}, where a given mode
  $\bk = 2\pi \tb{(m_x,m_y)}$ exhibits the fastest possible growth, is determined by
  the condition
  \begin{equation}
    \label{eq:kmaxgrowth}
    \frac{d}{d k}\frac{1}{\tb{\tau(k;\kappa,  T_\mathrm{eff})}} = 0
    \quad\Rightarrow\quad
    \tb{T_\mathrm{eff}(m_x,m_y;\kappa)} = \left(\frac{\capL}{2\pi}\right)^2 
    \frac{1 + (\capL/(2\pi))^2}{\left[ m_x^2+m_y^2 + (\capL/(2\pi))^2 \right]^2} .
  \end{equation}
  In particular, for the fundamental mode ($k=2\pi$) this defines the
  line $T_\mathrm{eff}=[1 + (2\pi/\capL)^2]^{-1}$ (red ``line F'' in
  Fig.~\ref{fig:phase}) above which the fundamental mode is the
  fastest growing one{\renewcommand{\baselinestretch}{1.0}\normalsize\footnote{The fundamental mode can actually still
    be the fastest growing one in the region even below this line but above
    the line corresponding to the next mode, $k=2\pi\sqrt{2}$. However, this is
    irrelevant for our purposes because the borders separating the
    different ``regimes of instability'' are not sharp anyhow.}}.
\end{enumerate}
It also proves useful to define the following loci:
\begin{enumerate}
\item The maximum value \tr{$1/\tau_{\mathrm{m}} =
    \underset{k}{\max}(1/\tau(k))$}  is given by,
\begin{equation}
  \label{eq:taum}
  \frac{1}{\tau_\mathrm{m}} = \left[ 1 -
    \sqrt{\frac{T_\mathrm{eff}}{1+(2\pi/\capL)^2}}\, \right]^2 ,
\end{equation}
which renders a line $T_{\mathrm{eff}}(\capL,\tau_{\mathrm{m}})$ with a parametric
dependence on $\tau_{\mathrm{m}}$. 
This time $\tau_{\mathrm{m}}$ is typically of order $1$, i.e., Jeans' time,
and it can become significantly larger only \tmg{close to the
  instability threshold}, $T_\mathrm{eff}\to 1$,
\tmg{and for a ``large system''}, $\capL\to\infty$; 
see in Fig.~\ref{fig:phase} the blue dashed line
corresponding to $\tau_\mathrm{m} = {10}$.
\item The line along which the characteristic time $\tau (k=2\pi)$ of
  the \tr{{\it f}undamental} mode has a given value $\tau_\mathrm{f}$ is given by
\begin{equation}
  \label{eq:tauf}
  \frac{1}{\tau_\mathrm{f}} = \frac{1-T_\mathrm{eff}}{1+(\capL/2\pi)^2} .
\end{equation}
\tmg{As illustrated in Fig.~\ref{fig:phase}, these lines run parallel
  to the line S and then become vertical at a sharp value
  \tr{$(\capL_0/2\pi)^2 = \tau_\mathrm{f}-1$}. In this sense, the notion
  of ``small vs.~large system'' is also quantified by the value of
  $\tau_\mathrm{f}-1$.}
\end{enumerate}
We can now summarize the overall picture in Fig.~\ref{fig:phase} as follows:
\begin{itemize}
\item In the region below the line S \tb{(stability limit)} and above
  the line F \tb{(\tr{the} fundamental mode is the fastest growing one)} there are in
  general many unstable modes \tmg{(thin blue lines)}, but the fastest
  growing one \tb{is} the fundamental mode (i.e., with the smallest $k$), so
  that the evolution 
  of the density field is dominated by features at the largest scales in real
  space (``collective collapse''). Furthermore, \tmg{in this region} the
  effect of a finite range 
  of the capillary attraction, i.e., $\capL\neq 0$, is irrelevant and the
  relative change in the growth rate between different modes is due to
  the effect of pressure (Eq.~(\ref{eq:tau})). This is the region of
  gravitational--like 
  collapse, which proceeds on a time scale of the order of Jeans' time
  (because \tmg{most part of} this region lies well below the line
  $\tau_\mathrm{m}={10}$ and \tmg{in the ``small system'' region,
    $\tau_\mathrm{f}\gtrsim 1$}).
\item In the region well below the red line F \tmg{and in the ``large
    system'' region, i.e., far to the right of the lines
    $\tau_\mathrm{f}\gtrsim 1$}, there are many unstable modes, but
  the fastest growing ones correspond to perturbations on small
  scales (compared to the size of the system), \tmg{which evolve on a
    time scale substantially faster than the fundamental mode}.
  Therefore, the evolution of the density field is dominated by the
  growth of \tmg{small--scale} perturbations of a certain
  preferred size \tmg{(``local collapse'')}.
  This is the region of spinodal--like instability, which also
  proceeds on a time scale of the order of Jeans' time, except in the
  corner region $\capL\to\infty$, $T_\mathrm{eff} - 1$ small (note the
  location of the line $\tau_\mathrm{m}={10}$). In this corner the
  evolution appears substantially slowed down on the scale of Jeans'
  time and it corresponds precisely to the usual scenario of spinodal
  decomposition in fluids \tmg{after quenching to a temperature below
    but not too far from the critical point} (note that the spinodal
  line is actually the line S).
\item The transition between \tmg{the two regimes discussed above
    is smooth and corresponds roughly} to the region below 
  the line F \tmg{but in the ``small system'' region, i.e., to the
    left of or around the lines $\tau_\mathrm{f}\gtrsim 1$. There the evolution
    consists of the simultaneous clustering at a rate of the order of
    Jeans' time of widely differing length scales, including both very
    small scales and the largest possible ones. This regime thus
    shares characteristic features of both the ``spinodal
    instability'' and the ``gravitational instability''. Which feature
    dominates the structure formation (global or local collapse)
    depends critically on the distribution of the initial amplitudes of
    the density perturbations.}
\end{itemize}
This picture is exemplified in Ref.~\cite{Bleibel:2011} with N--body
simulations in a finite box with periodic boundary conditions
corresponding to an initial dilute (ideal--gas) density with $1/K
\approx 8.75\times 10^{-3}$ fixed and values of $\capL$ ranging from $0.1$ up
to $100$; in Fig.~\ref{fig:phase} \tmg{the corresponding states are
  represented} by filled red squares lying on the iso-$K$ (thick black) line
$T_{\mathrm{eff}}(\capL)=K^{-2}\bigl[(2\pi)^2+\capL^2\bigr]$. \tr{In} these
simulations \tb{all initial state points were chosen below the stability line
  S, and 
  therefore the inherent mode instabilities lead to clustering. For
  the two lowest state points (I and II) with $\kappa<1$, perfect scaling of the
  time evolution of various geometric cluster measures with Jeans'
  time was observed, indicating the gravitational--like ``collective
  collapse''. The third state point (III) with $\kappa=1.85$ is located in
  the transition region and deviations from the above time scaling are
  noticeable.  For all other state points (IV--VII) with $\kappa \ge 5$, the
  dynamics became increasingly slow and the scaling with Jeans' time
  is completely lost. Together with the spatial information inferred from
  various snapshots (see Fig.~3 in Ref.~\cite{Bleibel:2011}), these state
  points are characteristic of the spinodal regime.  }

\subsection{\tmg{Collapse of a finite--sized, radially symmetric distribution}}
\label{sec:pert_summary}

\tmg{As compared to the case discussed in the previous subsection, the
  theoretical analysis of a collapsing finite--sized disk is
  \tg{complicated} by the fact that the unperturbed state is neither
  stationary nor spatially homogeneous. The theoretical description of such a
  system is facilitated by switching to Lagrangian
  coordinates.} The details of this treatment 
are presented in App.~\ref{app:pert}.  Here we summarize and discuss
the main results obtained from these calculations.

The Lagrangian trajectory field of volume elements is the mapping 
\begin{equation}
  \bx \longmapsto \br = \br_L(\bx,t) 
\end{equation}
such that $\br$ is defined physically as the position at time $t$ of that
volume element which was at position $\bx$ initially (i.e., at time $t
= 0$); $\bx$ is called the Lagrangian coordinate of the volume
element.
The trajectory field allows one to express any other field as function
of the Lagrangian coordinates $\bx$. In particular, the Lagrangian density field
is given by:
\begin{equation}
  n_L(\bx, t) := n( \br=\br_L(\bx, t), t) \;.
\end{equation}
Following the definitions and steps in App.~\ref{sec:pert}, the
evolution equations for $n$ and $w$ (Eqs.~(\ref{eq:w}) and
(\ref{eq:n})) can be put into their Lagrangian form
(see Eqs.~(\ref{eq:lagrg}\tr{)}, (\ref{eq:lagrw})\tr{,} and (\ref{eq:nevol})).  \tmg{Mass
  conservation is a built-in property of the Lagrangian scheme, so
  that it is not violated regardless of the approximations done in the
  Lagrangian equations. The \tr{subsequent} discussion is divided in two
  parts: the 
  radially symmetric evolution of the initial profile given by
  Eq.~(\ref{eq:n0}), and the (linearized) evolution of perturbations to
  this initial profile.}

\bigskip \tmg{\textbf{Radially symmetric evolution.} It is described by}
Lagrangian trajectories of the form $\br_L(\bx, t) = a(x, t)\, \bx$.
The formal solution for $a(x,t)$ is derived in
App.~\ref{sec:radsym}. \tmg{The theoretical analysis has been divided into
  steps of increasing difficulty:}

\begin{enumerate}
\item {\em Newtonian ($\kappa=0$), cold ($\,\Pi=0$) limit.} \tmg{The
    evolution equation for $n_L$ (Eq.~(\ref{eq:nevol})) can be solved
    exactly and the solution can be used to obtain $a(x,t)$ from
    Eq.~(\ref{eq:radiala}). In this manner one finds}
(see App.~\ref{sec:radsym_cold})
\begin{subequations}
  \label{eq:coldcollapse}
\begin{equation}
  a(x, t) =  \left\{
    \begin{array}[c]{cl}
      \displaystyle \sqrt{ 1 - t } , & x \leq 1 \\ 
      & \\
      \displaystyle \sqrt{ 1 - \frac{t}{x^2} } , & 1 < x 
    \end{array}
    \right.
\end{equation}
and
\begin{equation}
  n(\br,t) = \left\{
    \begin{array}[c]{cl}
      1/a^{2}(t) , & |\br| < a(t) \\
      & \\
      0 , & a(t) < |\br| \\
    \end{array} \right. .
\end{equation}
\end{subequations}
This is the so--called {\em cold collapse} solution, for which in the course
of the evolution the disk shrinks and the density inside the disk does not
depend on the radial position so that all
particles reach the center simultaneously. \tmg{This is the 2D analogue of
  the cosmological ``big crunch'' solution (i.e., a time--reversed ``big
  bang'' solution), where $a$ plays the role of the expansion factor
  and $da/dt$ that of the Hubble function.}

\begin{figure}
  \hfill\psfig{file=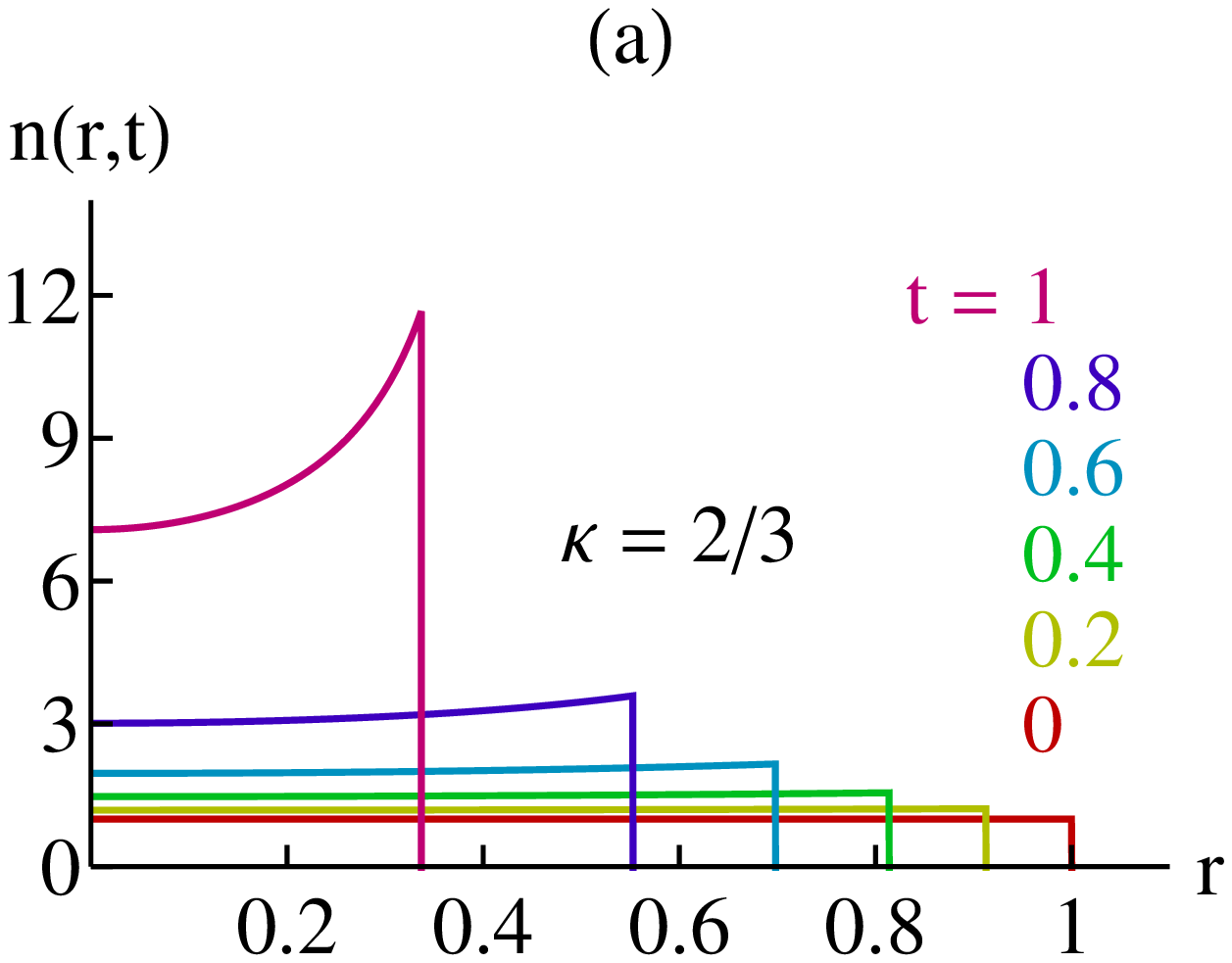,width=.4\textwidth}\hfill\psfig{file=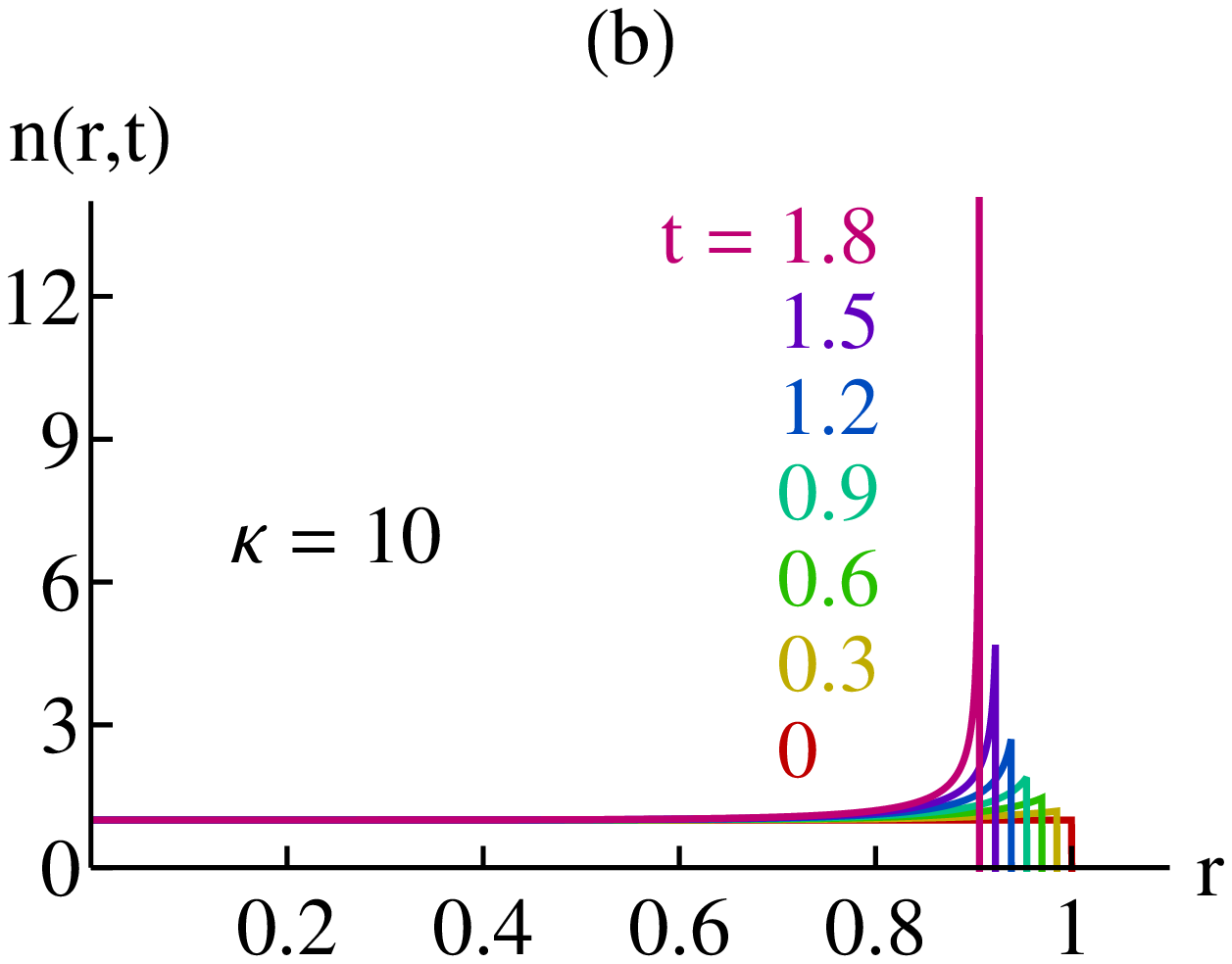,width=.4\textwidth}\hspace*{\fill}
  \caption{In the cold limit time evolution of the radial density profile for a
    disk computed perturbatively (see Eqs.~(\ref{eq:coldn}) and 
    (\ref{eq:colda})). (a) \tr{S}mall systems: for $\capL = 2/3$ the
    \tr{\textit{s}hock \textit{w}ave} 
    singularity appears at the radius $r_\mathrm{sw}\approx 0.17$ and at the time
    $t_\mathrm{sw}\approx 1.086$ (see Eq.~(\ref{eq:sw_pos})).
    (b) \tr{L}arge systems: for $\capL=10$ the shock wave singularity appears at
    the radius $r_\mathrm{sw}\approx 0.9$ and at the time $t_\mathrm{sw}\approx
      1.9$ (see Eq.~(\ref{eq:sw_pos2})).  
  }
  \label{fig:theorySW}
\end{figure}

\item {\em Finite screening ($\kappa \neq 0$), cold ($\,\Pi=0$) limit.}
  Analytical results can be obtained using a perturbative approach 
  \tmg{in the two limiting cases of ``small systems'' ($\capL\ll 1$) and of
    ``large systems'' ($\capL\gg 1$). This amounts to \korr{approximately}
    solving Eq.~(\ref{eq:nevol}) for $n_L$  by replacing the term
    proportional to $\capL^2$ by one corresponding to a reference solution. In
    the case $\capL\ll 1$, this is the cold--collapse solution given by
    Eq.~(\ref{eq:coldcollapse}).}
  The perturbative solutions for $a(x,t)$ and $n_L(x,t)$
  (Eqs.~(\ref{eq:coldn}) and (\ref{eq:colda})) result in a
  deformation of the density profile during the collapse (see
  Fig.~\ref{fig:theorySW}(a)). 
  In particular, at the rim of the disk a peak forms which becomes
  singular {\em before} all matter collapses in the center. To lowest
  order in $\kappa$, the radial position $r_\mathrm{sw}$ and \korr{the} time
  $t_\mathrm{sw} $ for the occurrence of this \tr{\textit{s}hock
    \textit{w}ave} are 
  given by 
  (Eqs. (\ref{eq:tswsmallkappa}) and (\ref{eq:rswsmallkappa}))
\begin{equation}
  \label{eq:sw_pos}
  \left. \begin{array}[c]{ccl}
      t_\mathrm{sw}  &\approx&  
      \displaystyle 1 + \frac{\capL^2}{8} \left[ \frac{1}{2} - 2 \gamma_\mathrm{e} - \ln\frac{\capL^2}{4} \right]  \\
      & & \\
      r_\mathrm{sw} &\approx& \displaystyle \frac{\capL}{4} 
  \end{array}
  \right\} ,
  \quad
  \capL\ll 1.
\end{equation}

\tmg{In the opposite limit of large system sizes, $\capL\gg 1$, the
  collapse of the disk is slowed down substantially because initially
  the density field is homogeneous almost everywhere (see
  Eq.~(\ref{eq:diff_large_kappa})). Thus one can use the \tr{initial
  condition as reference solution as if it \korr{was} stationary} in
  order to obtain an approximate solution of Eq.~(\ref{eq:nevolcold}).
  At the beginning, the dynamics is driven solely by the net capillary
  attraction near the density inhomogeneity at the rim and
  again a density singularity appears there (see
  Fig.~\ref{fig:theorySW}(b)). To lowest order in $\kappa^{-1}$, this
  occurs at a time and a position given by (see the text following
  Eqs. (\ref{eq:tsinglargekappa}) and (\ref{eq:coldalargekappa}))
\begin{equation}
  \label{eq:sw_pos2}
  \left. \begin{array}[c]{ccl}
      t_\mathrm{sw}  &\approx&  
      \displaystyle 2 - \frac{1}{\capL}  \\
      & & \\
      r_\mathrm{sw} &\approx& \displaystyle 1 - \frac{1}{\capL} 
  \end{array}
  \right\} ,
  \quad
  \capL\gg 1.
\end{equation}
After this, the position of the singularity is shifted by the collapse
of the disk, which occurs on a much longer time scale \tr{which is} at least
$\gtrsim 1/\capL$ \tr{(}see Eq.~(\ref{eq:coldalargekappa})).}

\tmg{Thus, the formation of a singularity is a generic feature of the
  cold limit. The time evolution is formally undefined beyond its
  appearance, but actually the singularity is regularized by
  the effect of a nonzero compressibility caused by corrections to the
  cold limit.}

 \item {\em Effects of nonzero $\Pi$.} 
At nonzero temperatures, the pressure term in the evolution equation for $n$
(Eq.~(\ref{eq:n}) or, equivalently, 
the Lagrangian form in Eq.~(\ref{eq:nevol})) can become important. 
\tmg{If $\Pi$ is sufficiently large (formally the limit of high temperature, $T\to\infty$), pressure is
  the dominant term in the evolution equation and the initial density
  distribution \korr{ends up with two--dimensional evaporation} instead of
  \korr{a} collaps\korr{e}. This is the analogue
  of the region $T_\mathrm{eff}>1$ in Fig.~\ref{fig:phase}. We are
  more interested, however, in the case that the disk collapses and
  how a nonzero, but still sufficiently \korr{low} temperature $T$ affects the
  evolution.}
In App.~\ref{sec:Newthot} we show  
that a nonzero pressure quickly smoothens any finite steps in the initial
density distribution. \tmg{The analytical study of the effect of pressure on
  the density singularity, which forms as a divergence for $\Pi=0$ and
  $\capL\neq 0$}\tr{,} is more difficult and we do not have explicit analytical
results for this behavior, \tmg{but it can be expected that}  
the pressure will regularize also this singularity. 
This is clearly visible
in our \tmg{numerical} solutions and in the simulated profiles (see
Sec.~\ref{sec:simresults} below). Thus the collapse 
will proceed beyond the time $ t_\mathrm{sw} $ and the regularized peak at the
rim of the disk is shifted 
to the center in the manner of a {\em shock wave}.
\end{enumerate}

\bigskip \tmg{\textbf{Perturbations of radial symmetry.} } 
This issue has been studied in App.~\ref{sec:nonrad}. For the Newtonian,
cold limit \tmg{one can solve the evolution equation for the
  Lagrangian density field \emph{exactly} (see Eq.~(\ref{eq:frag})).
  There is enhanced clustering localized at initially overdense
  regions and depletion at underdense regions. In astrophysics this is
  called the ``fragmentation instability''.  For an initially small
  local overdensity, $0<\delta n_0\ll 1$, the local time of collapse
  (signaled as a density \tr{\textit{s}ingularity} ) is still of the order of
  Jeans' 
  time, $t_\mathrm{s} \approx 1 - \delta n_0$.} 
Again, it is difficult to treat the case of \korr{nonzero} $T$ and $\kappa$.
However, for density perturbations localized near the center of the
disk \tmg{such that the effect of the boundary can be neglected}, the
calculations in App.~\ref{sec:nonrad} show that \tmg{
  the linear stability analysis of a \tr{time--independent}, homogeneous state
  as discussed in Subsec.~\ref{sec:hom} is useful also for this kind of
  perturbations. Particularly relevant observations are that the
  shortest temporal scale for the growth of perturbations is in general
  Jeans' time (see the line denoted as $\tau_\mathrm{m}$ in
  Fig.~\ref{fig:phase}) and that during disk collapse} further fragmentation
of the inhomogeneities is counteracted by a nonzero pressure.

\bigskip

\tmg{Although Fig.~\ref{fig:phase} \korr{derives from} the linear
  stability analysis of a \tr{stationary} and homogeneous state, it is useful
  to summarize and rationalize also the picture which emerges from the
  theoretical analysis of the collapsing disk, with $T_\mathrm{eff}$
  and $\capL$ referring to the initial disk state. Thus, the disk
  collapses if $T_\mathrm{eff}$ is well below $1$. A generic feature
  of the disk--collapse is the formation of a singularity, eventually
  regularized by pressure, at the outer rim at a time slightly larger
  than Jeans' time. As an inbound shock wave this feature can dominate
  the evolution only in an intermediate transition region where
  $\capL$ is of \tr{the} order \tr{of} one and the time scale of the overall
  collapse is comparable to Jeans' time (see the lines denoted as
  $\tau_\mathrm{f}$ in Fig.~\ref{fig:phase}). If $\capL$ is very
  small, the gravitational--like global collapse is so fast that the
  formation of the singularity is practically unobservable. If $\capL$
  is very large, the evolution is dominated by spinodal--like growth,
  on Jeans' time scale, of density perturbations with large
  wavelengths inside the disk, while the collapse of the disk remains
  frozen and with it the singularity at the outer rim. This picture is
  confirmed by numerical calculations, as explained in the next
  sections.}

\section{Brownian dynamics and particle based DDFT}
\label{sect:2DDDFT}

The Brownian dynamics simulation introduced in Ref.~\cite{Bleibel:2011} relies
on the \tr{particle--mesh} (PM) method known from cosmological simulations. We
first briefly recall the basic concept of the PM method, which 
provides the foundation of an easily applicable extension corresponding to a
solution of a \tb{two--dimensional \korr{(2D)} dynamical density functional
  theory (DDFT)}.

\tb{Within simulation,} the straightforward way to implement \tb{colloidal}
dynamics based on a pairwise additive  
potential \tr{such} as \tb{the capillary potential of}  Eq.~(\ref{eq:2part})
\tb{consists of} summing  all 
forces from all 
possible pairs of colloids.
However, as explained after Eq.~(\ref{eq:2part}), the potential in
Eq.~(\ref{eq:2part}) is long--ranged and
non--integrable in 2D \tmg{in a system of size $L<\lambda$}.
In using periodic boundary conditions, colloidal particles
from periodic images will therefore contribute significantly to the total
force acting on a single colloid. In order to obtain the correct value of the
force 
experienced by a particle, the sum of the forces has to be extended over many
periodic images. Alternatively, \tb{here} we use the particle--mesh
method, in order to
exploit the underlying mean field character of Eq.~(\ref{eq:YL}). It connects
the average interfacial deformation $U(\vect{r})$ with the average mean number
density
$\varrho(\vect{r})$. Provided one has access to the mean density, 
e.g., discretized on a grid or mesh, one can solve Eq.~(\ref{eq:YL}) via 
Fourier transformation. The resulting mean interfacial deformation on
the grid enables one to compute the capillary forces acting on the colloidal
particles by means of interpolation, \tmg{whereby the effect of periodic images is taken into account properly by the Fourier transform}.      

In order to simulate colloidal particles trapped at a fluid interface,
one also has to consider the effect of the thermal heat bath provided by the
fluid, as well as the nonzero size of the particles. \tb{In continuum theory,}
both aspects are taken into account by the
presence of the \tb{(local)} term $\propto \nabla p(\varrho)$ in
Eq.~(\ref{eq:cont}).   
Within Brownian dynamics (BD) simulations, \tb{the nonzero particle size is accounted for
by 
short--ranged repulsive forces between the colloids. We use
the cut off and shifted repulsive part of the
Lennard--Jones--potential (\tmg{the so--called soft WCA potential,} as used in
Ref.~\cite{Bleibel:2011}).
Temperature enters  by adding a
stochastic noise term to the equations of motion of the particles.} Thus,
within BD one
integrates the corresponding Langevin equation.

\tb{Thus, our BD simulations} incorporate two different methods for the
computation of the forces: the \tr{particle--mesh} method for the long--ranged
capillary forces and the direct summation of short--ranged repulsive forces. 
It would be even more convenient, if the
particle--mesh method could be extended towards the short--ranged forces. Such
an 
approach has been used in Ref.~\cite{Gnedin:1998} for describing the very dilute
intergalactic medium.
\tmg{The same idea can be employed for Eq.~(\ref{eq:nwithmu}) because}
the forces $\propto \nabla (\mu-w)$ can easily be
calculated for the discretized density grid. 
Since \tr{the}  
fluctuations\tr{,} giving rise to the noise present in BD simulations\tr{,}
are now also 
incorporated \tr{within a} mean field \tr{approximation},  
this method \tb{describes} the dynamical evolution of the discretized
density \tb{according to an initial condition for the density field set by the 
distribution of particles at time zero. The evolution of the ensemble--averaged
density field is obtained by averaging over initial conditions, i.e., the
initial distributions 
of particles. In that manner, the \tmg{obtained averaged evolution}
corresponds to the solution of the 
2D DDFT defined by the evolution equation (\ref{eq:ddft}). } 
We call it particle based, because in every
time step one discretizes the density, i.e., one assigns all particles to
points on 
the grid and calculates the forces $\nabla (\mu-w)$ on the grid. Then, instead
of integrating Eq. (\ref{eq:nwithmu}) one interpolates these forces
back to the positions of the particles, which they occupied before they were
assigned to grid points, and integrates their deterministic
equations of motion. The next step, which is to discretize the density at time
$t+\Delta t$\tr{,} completes the evolution of the density as the basic
quantity of our analysis. As has been
pointed out in Ref.~\cite{Gnedin:1998}, in comparison with the standard
\tr{particle--mesh} method, this discretization and integration
scheme comes at the cost of an increased numerical noise at small scales of
the order of the grid-spacing. However, for any physical observable we have
studied so far, this increased 
numerical noise turned out to be not relevant.

\section{Results from simulations}
\label{sec:simresults}

\tb{Using BD simulations and 2D DDFT as described in the previous section,
we explore the \tr{``dynamical phase diagram''} (Fig.~\ref{fig:phase}) for the
initial 
condition of a finite--sized patch of particles,
arranged in a circular disk. In order to locate in the
$T_\mathrm{eff}$--$\kappa$ plane of Fig.~\ref{fig:phase} the kind of evolution
of a certain configuration, the value 
of the effective temperature (Eq.~(\ref{eq:Teff})) is determined by the
initial conditions inside the disk.
To this end we recall the central result from the linear stability analysis 
 in \tr{Subsect.} \ref{sec:hom} and the perturbative analysis for the disk
 evolution 
 in \tr{Subsect.} \ref{sec:pert_summary}: When following an isothermal
 line{\renewcommand{\baselinestretch}{1.0}\normalsize\footnote{\label{foot_5}\tb{We recall that the physical temperature $T$
     enters  
 Jeans' length via $K^{-1}=(d\Pi/dn)^{1/2}$. Thus for the repulsive
 interaction taken to be of the hard 
 core type, one has $K^{-1}\propto T^{1/2}$. Since the effective temperature
 $T_\mathrm{eff}$ used in Fig.~\ref{fig:phase} is given by
 $T_\mathrm{eff}=[(2\pi)^2+\kappa^2]/K^2$ (Eq.~(\ref{eq:Teff})), in the
 double-logarithmic $T_\mathrm{eff}$--$\kappa$ plane of Fig.~\ref{fig:phase}  
 an isotherm is a line which is horizontal for small $\kappa$ and crosses over
 to a line  with slope 2 at large $\kappa$ due to $T_\mathrm{eff}\propto
 \kappa^2$.}}}  from large capillary lengths $\lambda$ (i.e., small
cutoff parameters $\kappa=L_0/\lambda$) down to \tr{$\lambda=2R$}, i.e., the value at
which the range of the interaction  
equals the particle diameter}, one first observes a collective
collapse, followed by a regime where shock waves become visible and a spinodal
decomposition stage for rather short--ranged attractions, i.e., $\lambda/L_0 \ll
1$ { or $\kappa \gg 1$}.      

We fix the temperature of the system by choosing a disk of radius \tr{$L_0=180
  R$} 
  and setting the number of particles with radius \tr{$R$} to $N=1804$ so that
  $\varrho_0\simeq 574/L_0^2$. The
  associated \tr{dimensionless} Jeans' length{\renewcommand{\baselinestretch}{1.0}\normalsize\footnote{As discussed in
    Ref.~\cite{Bleibel:2011}, we employ an ideal gas equation of state for the
    initial condition to \tr{determine} Jeans' length according to
    $K^{-1}\approx 
    \sqrt{\gamma \kt/(f^2\varrho_0)}$ with $f^2/(2\pi\gamma)=0.89 \,\kt$ (see
    Eq.~(\ref{eq:2part}); \tr{this numerical value corresponds to the choice
    $R=10$ $\mu$m~\cite{Oettel:2008}) so that $K^{-1}\simeq 0.42
    \varrho_0^{-1/2}$}.}} 
  \tr{in units of $L_0$} is $K^{-1}=0.018$\tr{,}  corresponding to an
  isotherm\tr{\footref{foot_5}} 
 containing the filled circles (a)--(c) in
  Fig.~\ref{fig:phase}. For this isotherm we have carried out simulations at
  $\kappa=0.67$ (collective collapse regime\tg{, $\tau_{\rm f}=1.01$}),
  $\kappa=4.0$ (shock wave regime\tg{, $\tau_{\rm f}=1.45$})\tr{,} and
  $\kappa=40.0$ 
  (spinodal regime\tg{, $\tau_{\rm f}=41.5$}).  
\label{sect:result}
\begin{figure*}[ht!]
  \begin{minipage}{0.32\linewidth}
    \epsfig{file=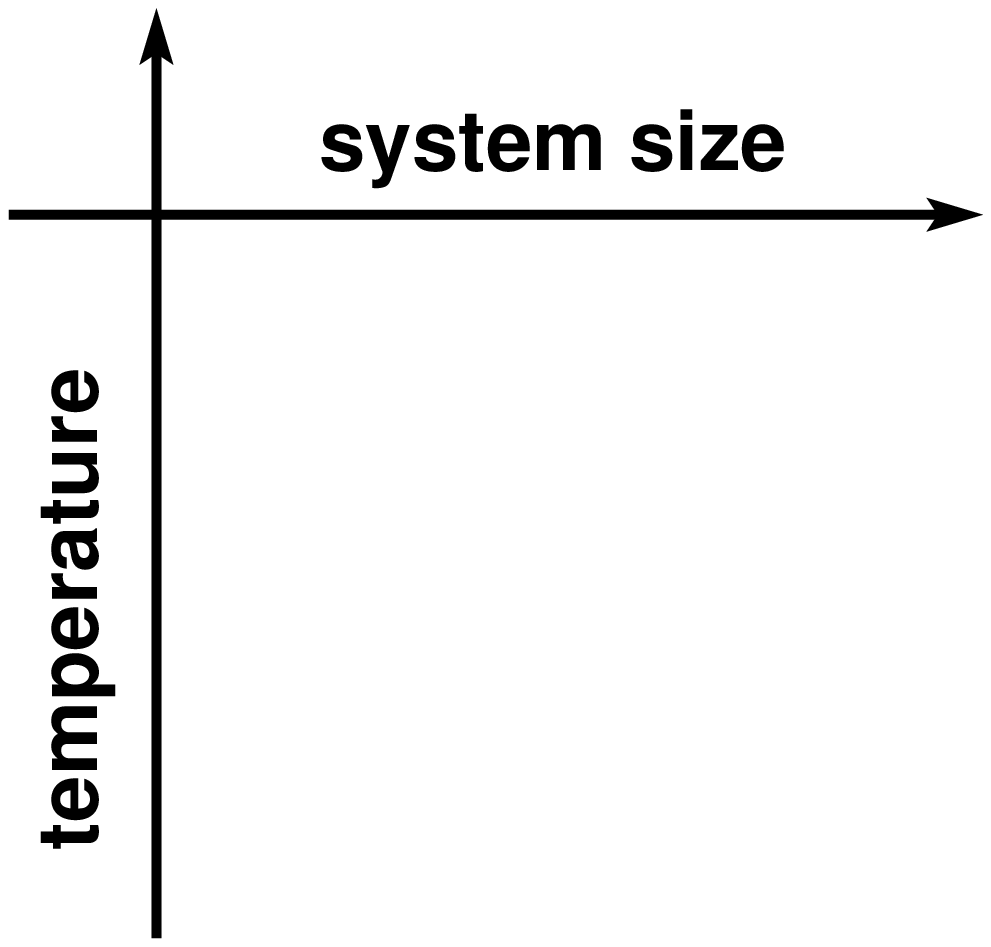,width=0.75\linewidth}
  \end{minipage}
  \begin{minipage}{0.32\linewidth}
    \epsfig{file=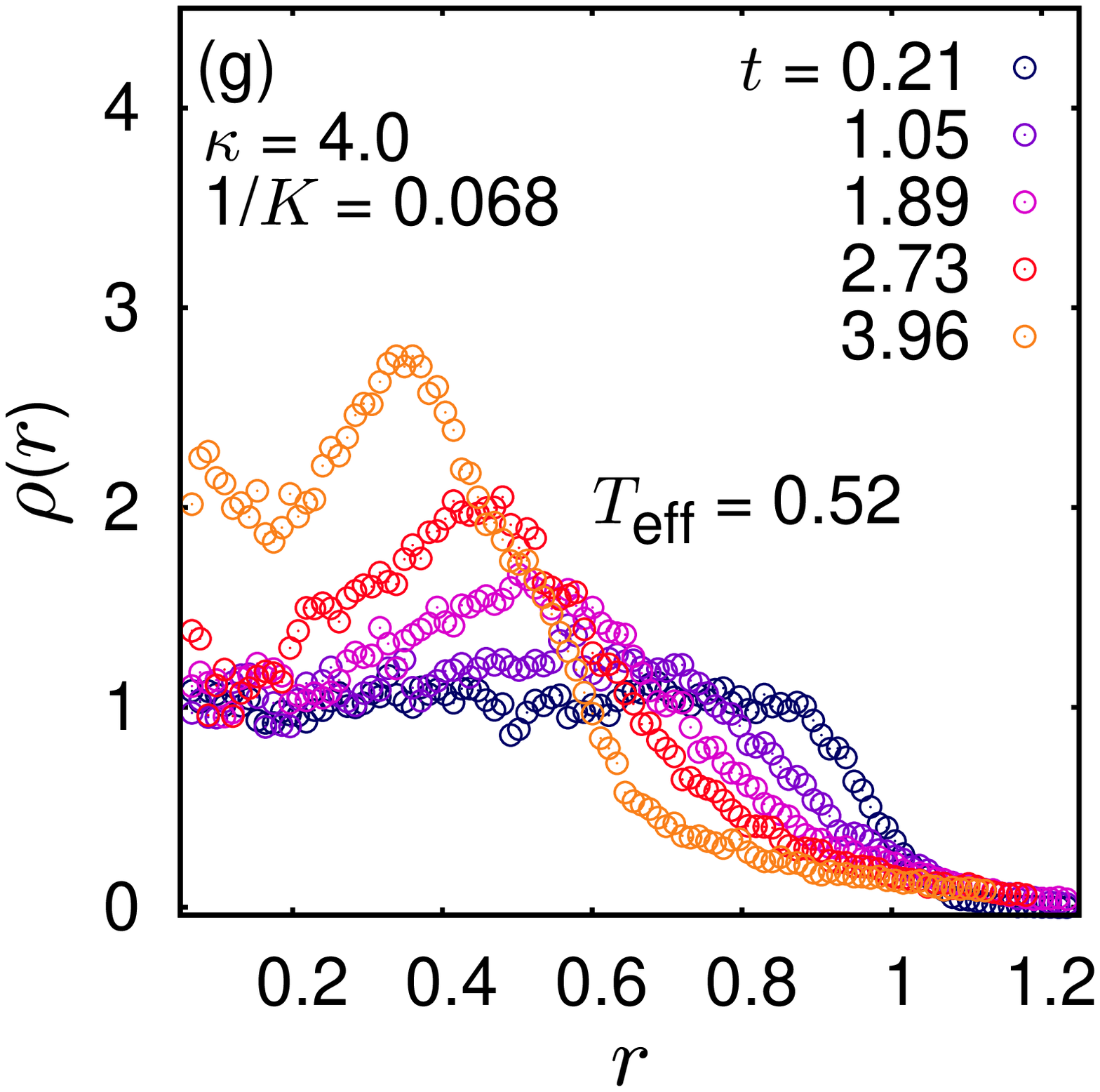,width=\linewidth}
  \end{minipage}
  \begin{minipage}{0.32\linewidth}
    \epsfig{file=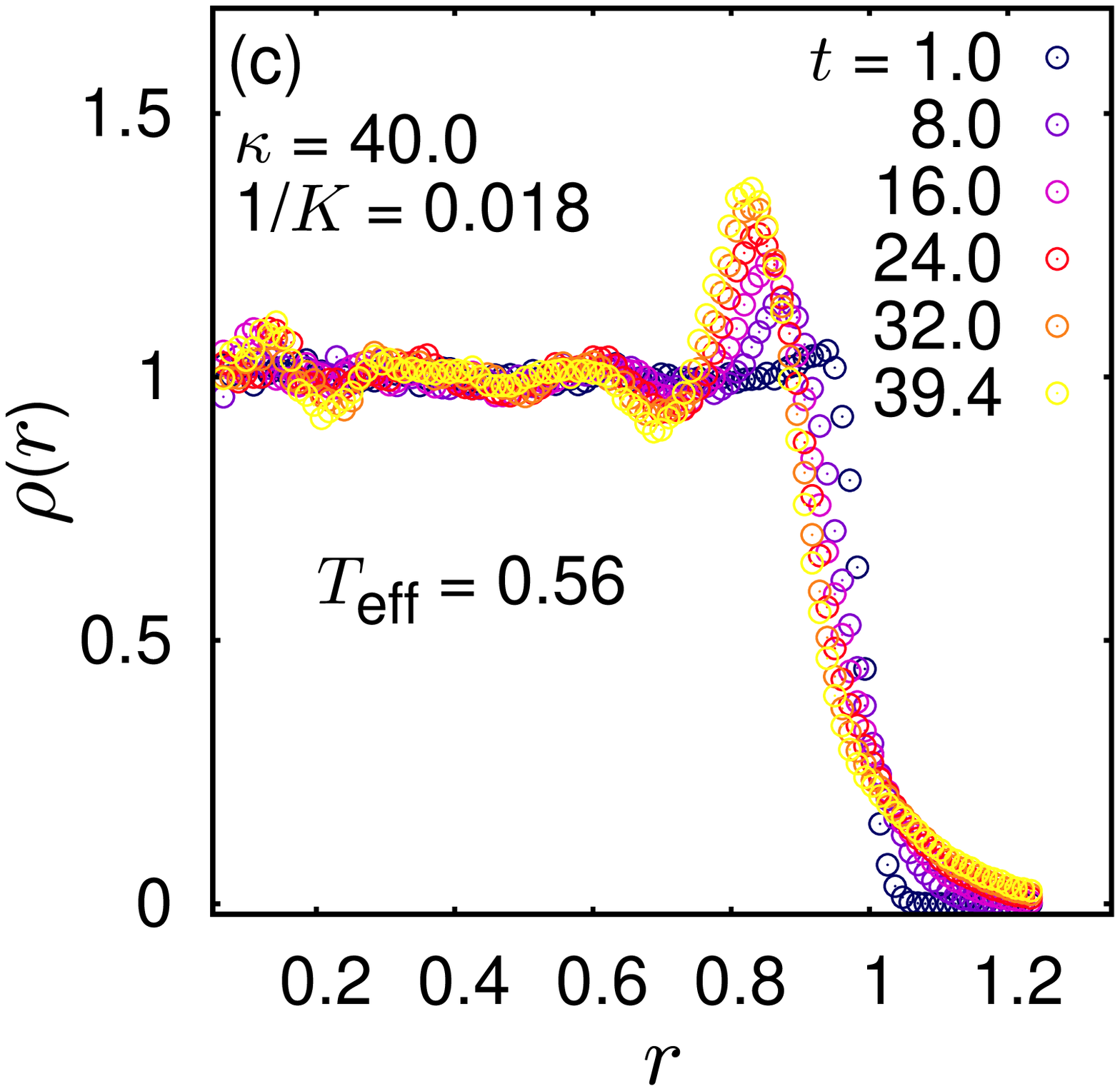,width=\linewidth}
  \end{minipage}
  \begin{minipage}{0.32\linewidth}
    \epsfig{file=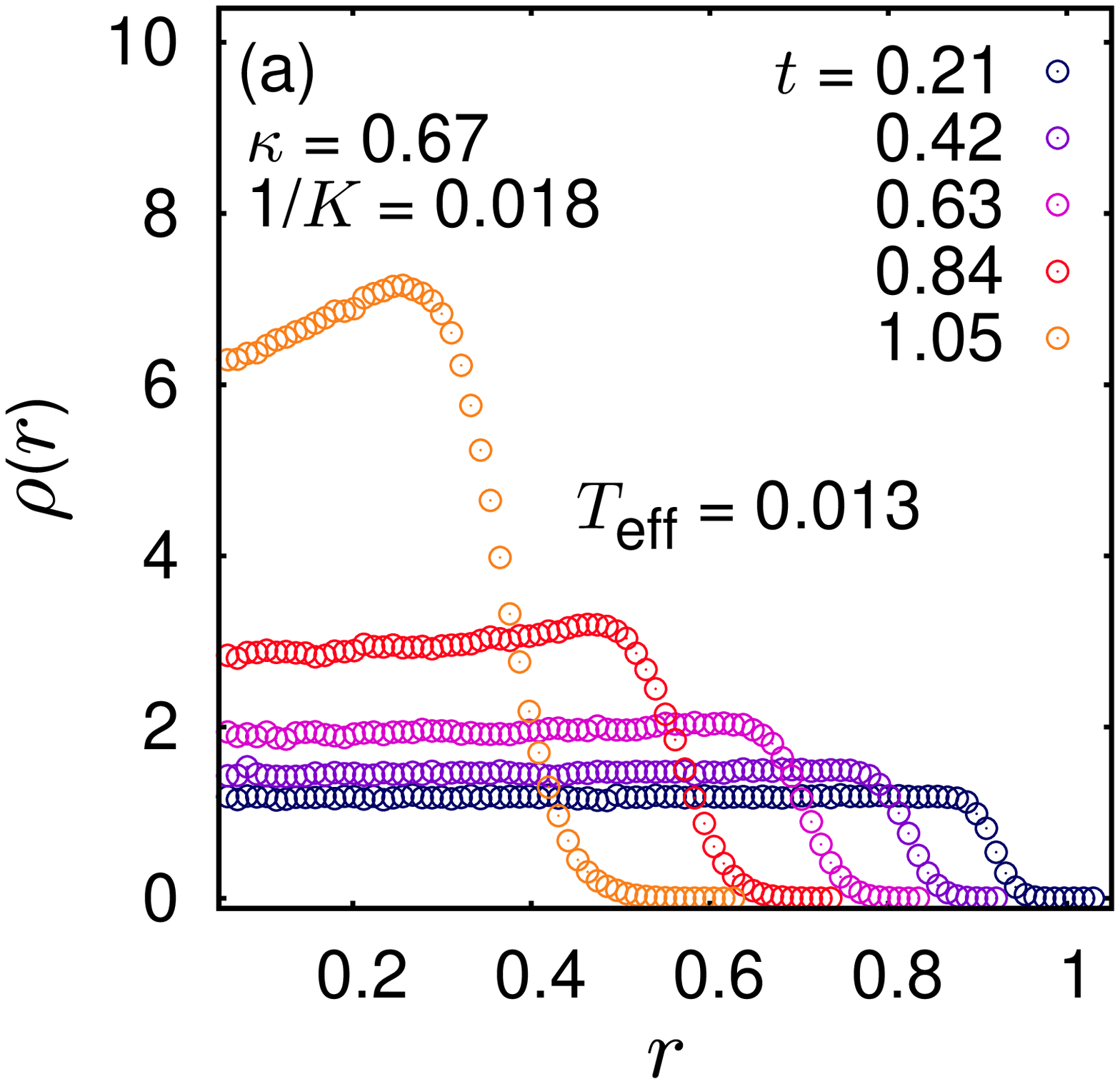,width=\linewidth}
  \end{minipage}
  \begin{minipage}{0.32\linewidth}
    \epsfig{file=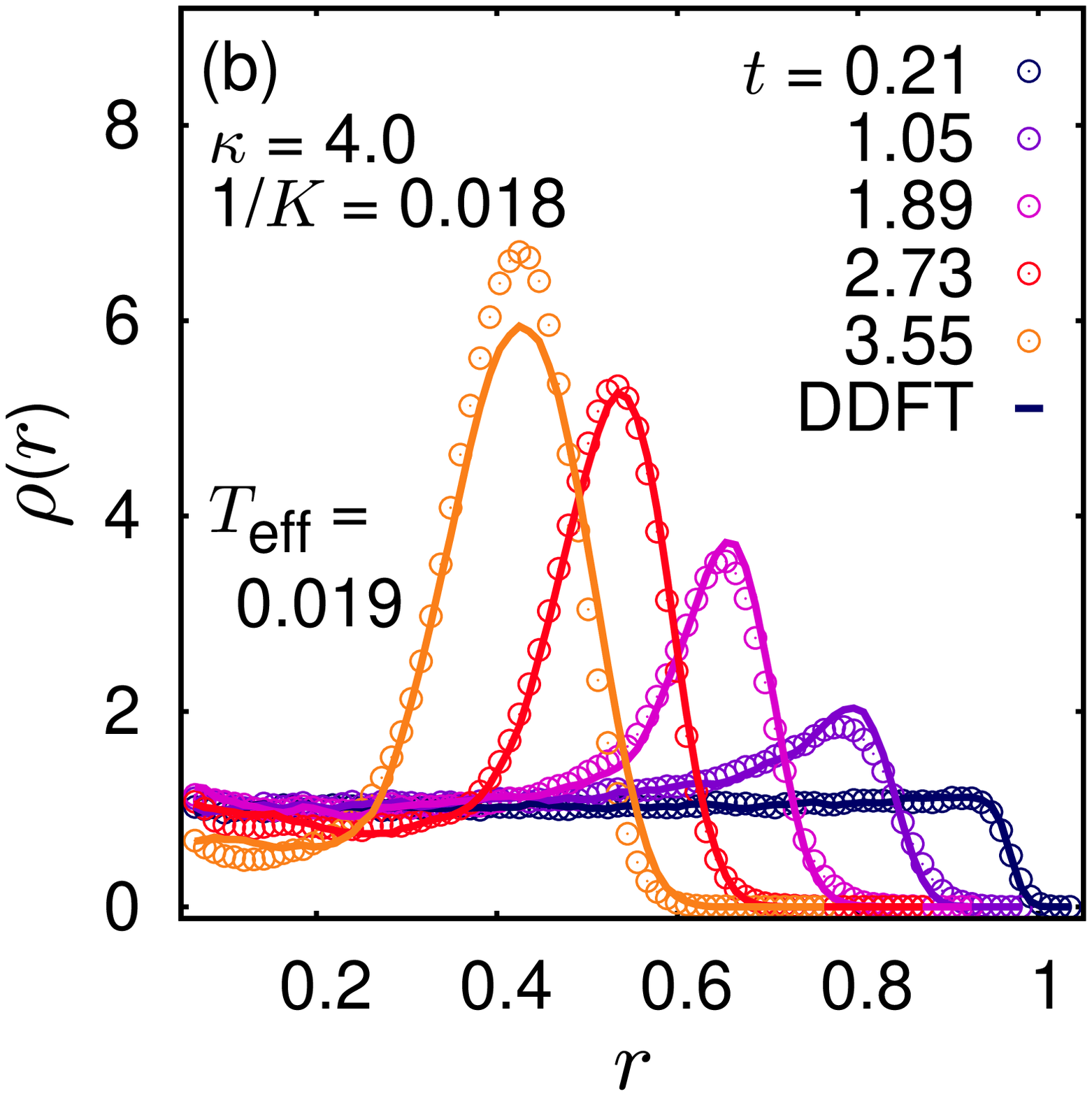,width=\linewidth}
  \end{minipage}
  \begin{minipage}{0.32\linewidth}
    \epsfig{file=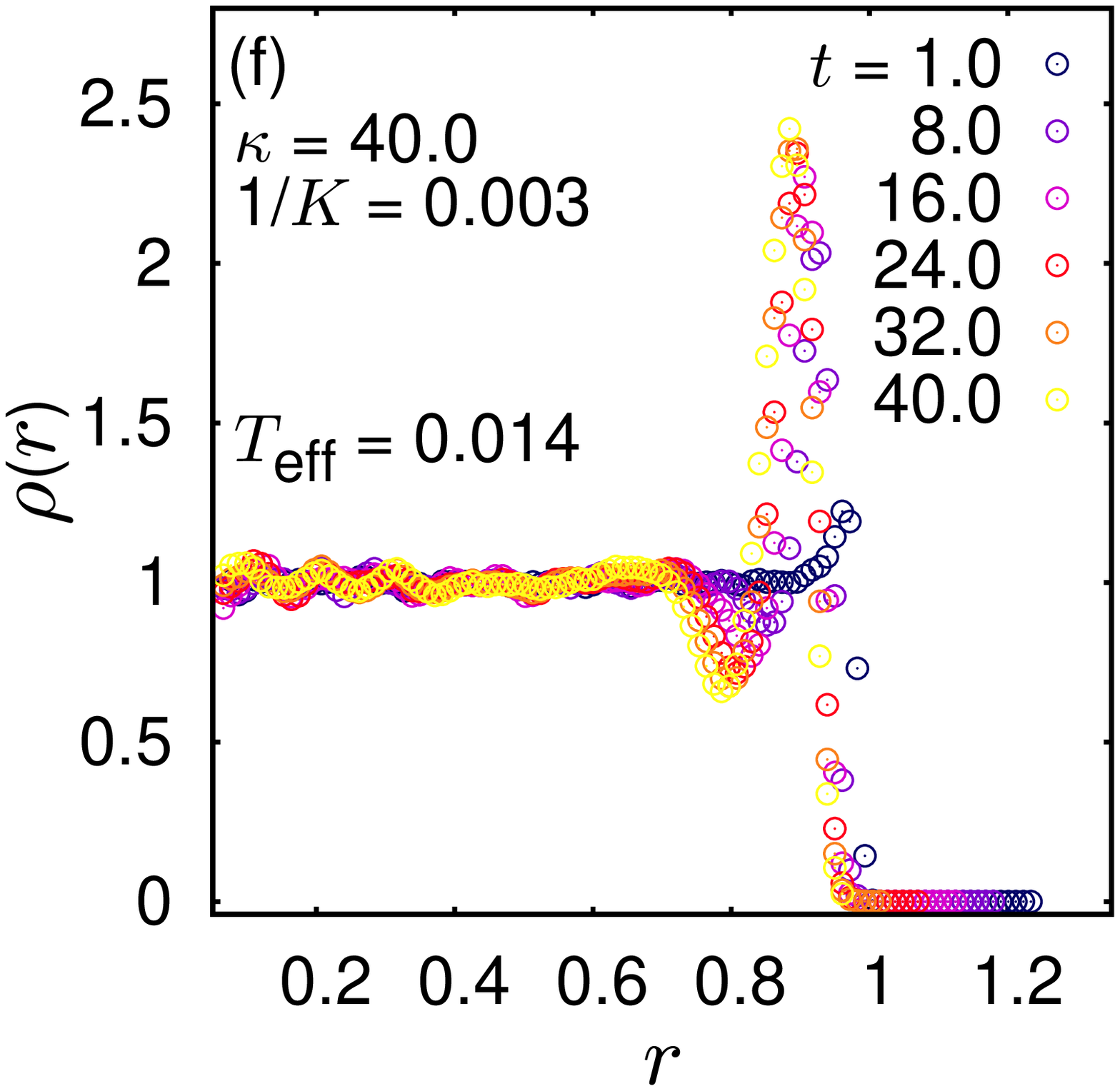,width=\linewidth}
  \end{minipage}
  \begin{minipage}{0.32\linewidth}
    \epsfig{file=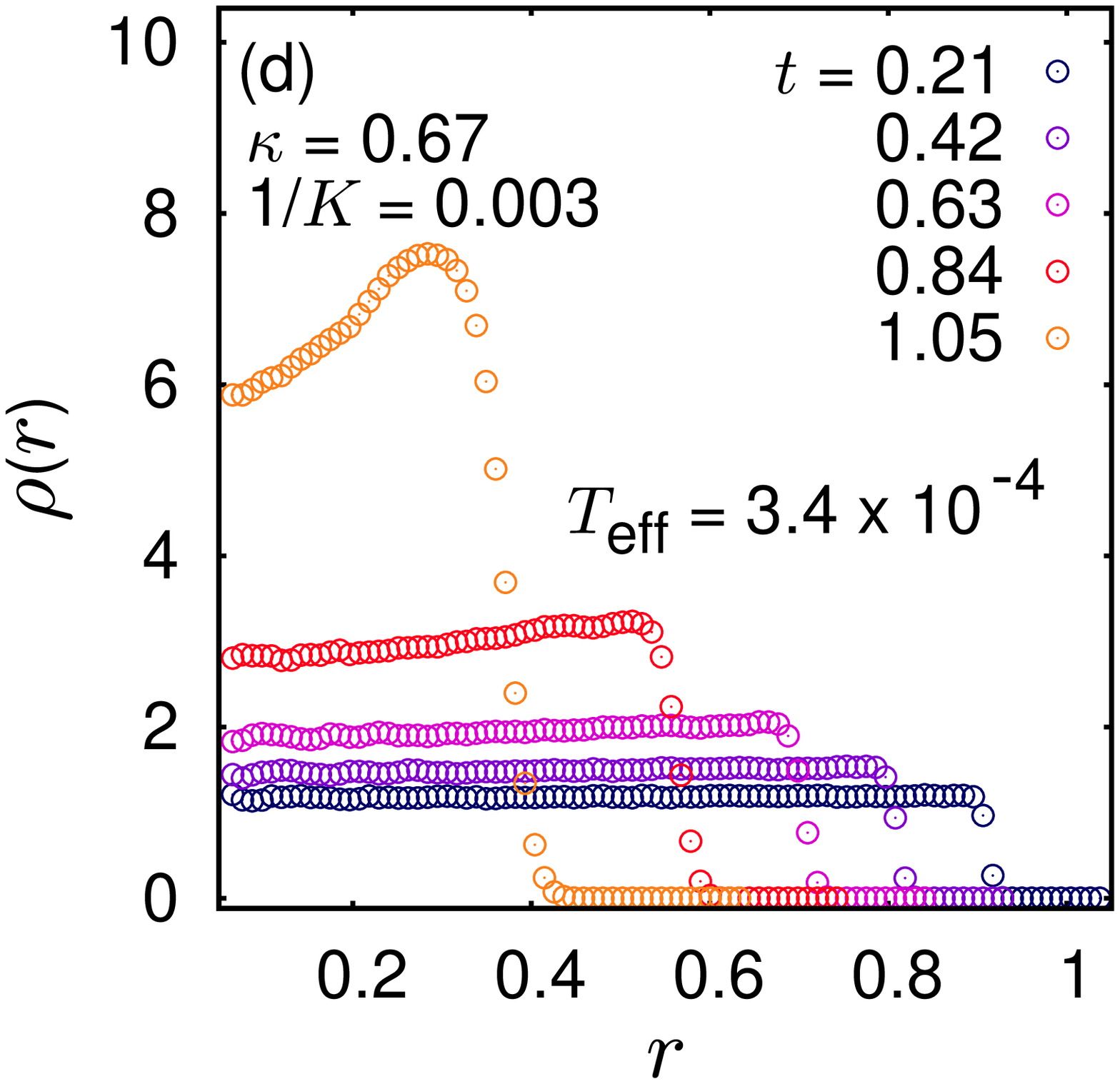,width=\linewidth}
  \end{minipage}
  \begin{minipage}{0.32\linewidth}
    \epsfig{file=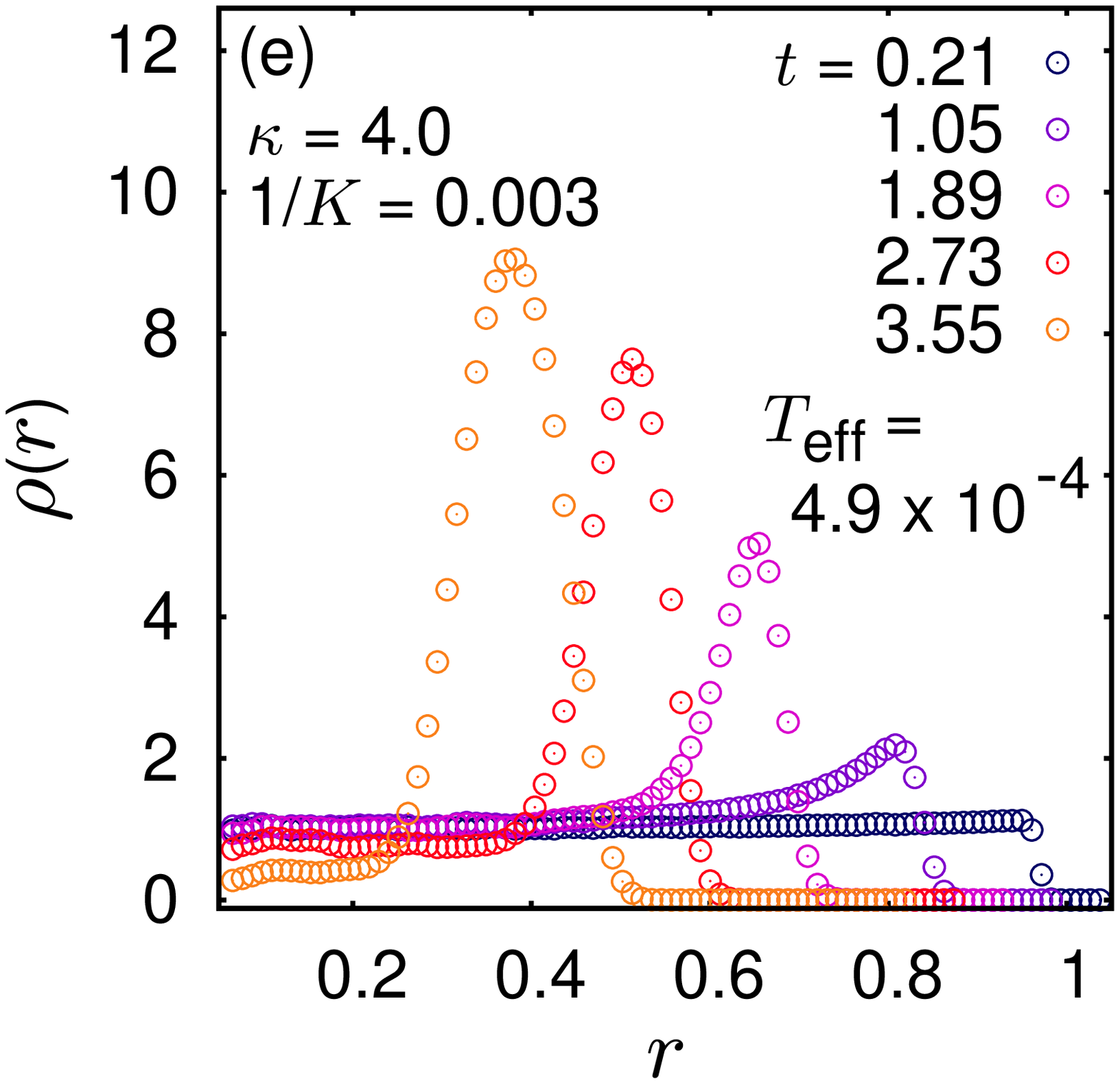,width=\linewidth}
  \end{minipage}
  \begin{minipage}{0.32\linewidth}
    \epsfig{file=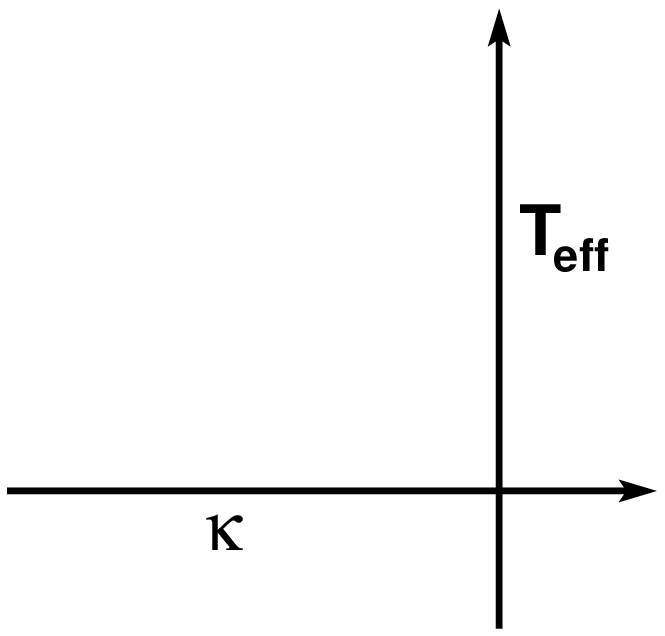,width=0.75\linewidth}
  \end{minipage}
  \renewcommand{\baselinestretch}{0.92}\normalsize
  \caption{\label{fig:radial1} Evolution of the radial
    density profiles \tg{corresponding to the points (a)--(g) in the
      \tr{``dynamical phase diagram''} in Fig.~\ref{fig:phase}. As indicated
      by the arrows, the 
    reduced system size $\kappa=L_0/\lambda$ increases from left to right, the
    effective temperature $T_{\rm eff}$ increases from the bottom panels to
    the top ones.}    
    \tg{Panels (a)--(c): Density} profiles for the  
    effective temperatures $T_{\rm eff}=1.3 \times 10^{-2}$ \tg{(a)}, $T_{\rm
        eff}=1.9 \times 10^{-2}$ \tg{(b)}, and $T_{\rm eff}=0.56$ \tg{(c)} 
      (corresponding
    Jeans' length: $K^{-1}=0.018$) and for three values for the cutoff
    parameter $\kappa=L_0/\lambda$. Panel (a): $\kappa=0.67$, results of
      Brownian dynamics simulations (averaged over 120 runs), no clear signal
      of a shock wave is visible. Panel (b): $\kappa=4.0$, comparison
    between 2D DDFT (colored lines, averaged  over 120 runs) and the
    Brownian dynamics 
    simulations (symbols). Panel (c):  $\kappa=40.0$, results of Brownian
    dynamics simulations (averaged over 500 runs). 
    In contrast to (b), in (c)
    transient peaks resulting from clustering close to the center of the
    collapsing disk but separated from the shock wave become visible.
    \tg{Panels (d)--(f): radial density profiles at lower effective
      temperatures $T_{\rm eff}=3.4 \times 10^{-4}$ (d), $T_{\rm eff}=4.9 \times
      10^{-4}$ (e), and $T_{\rm eff}=1.4 \times 10^{-2}$ (f) (corresponding
    Jeans' length: $K^{-1}=0.003$) for $\kappa=0.67$, $\kappa=4.0$, and
    $\kappa=40.0$\tr{,} respectively (Brownian dynamics, averaged over 500
    runs).} 
    \tg{Panel (g): Same as above, at an effective temperature of
    $T_{\rm eff}=0.52$ (corresponding Jeans' length:
    $K^{-1}=0.068$) for $\kappa=4.0$ (Brownian dynamics, averaged over 500
    runs). An inbound traveling shock wave is observed clearly for $\tau_{\rm
      f}\gtrsim 1$, which is realized by the systems (b) and (e) ($\tau_{\rm
      f} = 1.45$).} 
    } 
\end{figure*} 
 Figure~\ref{fig:radial1} shows the evolution of the radial density profiles
  for these three values of $\capL$ \tg{at various temperatures
    $T_{\rm eff}$. (In Fig.~\ref{fig:radial1} the panels are ordered in the
    plane in correspondence to the locations of the 
    points (a)--(f) in the \tr{``dynamical phase diagram''}
    in Fig.~\ref{fig:phase}).} The Brownian dynamics 
  simulations  are in line with the particle based DDFT (see
  Fig.~\ref{fig:radial1}(b)). For 
    $\kappa=0.67$ (point (a)) the evolution is dominated by the collective
    inbound motion, resembling almost the cold collapse scenario. For
  $\kappa=4.0$ \tb{(point (b))} a pronounced 
  shock wave is visible. The time scale for the collapse is stretched by a
  factor of ca. $4$ \tb{compared with the time scale for the cold collapse}. 
  The fundamental mode corresponding to the largest
  scales still grows \tr{rather} fast in this scenario\tg{:
    $\tau_\mathrm{f}=1.45 \gtrsim 1$\tr{;} therefore, this kind of evolution
    corresponds to the transition 
    region between collective collapse and spinodal decomposition}. 
 \tb{In contrast, for  the}
  simulations with the rather large cutoff parameter $\kappa=40$ \tb{(point
    (c))}, associated 
  with a data point in the regime of spinodal decomposition, \tr{quite a}
  different picture emerges.
{As has been anticipated in \tr{Subsect.}~\ref{sect:theory}, this limit
  corresponds to the diffusive regime with a \tb{comparatively} short--ranged
  attractive force 
  present (see Eq.~(\ref{eq:diff_large_kappa})). Inspecting the \tr{``dynamical
  phase diagram''} 
  (Fig.~\ref{fig:phase}), one finds that \tb{point (c) is located in} 
   the top right corner. The amplitude for the growth of the
  fluctuations is rather small; for the fundamental mode it approaches
  zero. Since point \tb{(c)} in the phase diagram is well above the line
  corresponding to \tg{$\tau_{\rm f}=41.5$}, the characteristic time of the
  fundamental 
  mode will be \tg{even} larger, leading to a very slow inbound motion of the
  density 
  peak developing at the outer rim. In order to estimate the characteristic
  time for the collapse in this case, we rescale the density $n$ in
  Eq.~(\ref{eq:diff_large_kappa}) according to $n^{\,\prime}=\sqrt{2} \kappa
  n$, which 
  leads for a dilute system ($\Pi(n) \ll 1$) to a rescaled characteristic time
  $\mathcal{T}^{\,\prime} = \sqrt{2} \kappa \,\mathcal{T} \approx 57\,
  \mathcal{T}$. The collapse towards a close packed patch is therefore
  expected to occur at
  times $t > \mathcal{T}^{\,\prime}$. This is in line with the observed slow
  inbound 
  motion of the density peak. Additionally, at least one secondary peak in the
  inner part becomes visible for large times. These secondary peaks are much
  more pronounced 
  if the effective temperature is lowered (see \tg{Fig.~\ref{fig:radial1},
  panel (f)}).}   

  Next we discuss an isotherm at a lower temperature which results in an
  effective temperature $T_\mathrm{eff}$ about two orders of magnitude 
  \korr{lower} than the previous one ($N=3844$, $K^{-1}=0.003$), see the black,
  filled triangles 
  (d), (e)\tr{,} and (f) in Fig.~\ref{fig:phase}. (To this end the physical
  temperature does not need 
  to be changed by two orders of magnitude. Such a shift can also be achieved
  by changing the initial number density, or the capillary strength $f$ via
  the particle size, because these quantities enter \tr{into} the reduced
  pressure $\Pi$ (Eq.~(\ref{eq:dimless})) and thus also \tr{into} Jeans'
  length and \tr{into} $T_\mathrm{eff}$.)     
 In Fig.~\ref{fig:radial1} \tr{p}anels (d), (e)\tr{,} and (f) show the radial
evolution of the density, \tb{as in (a)--(c), for 
   $\kappa=0.67$ (point (d)), $\kappa=4.0$ (point (e)), and $\kappa=40.0$
     (point (f)).}  
  In the latter two cases one observes peaks which are somewhat narrower than
  in (b) and (c), in line
  with the 
  reduced pressure at lower temperatures. However, the qualitative behavior of
  the inbound motion is different. For $\kappa=4.0$ the collapse is
  accelerated \tb{compared with point (b) (Fig.~\ref{fig:radial1}(b)
    corresponds to the 
   same $\kappa$ but to a higher $T_\mathrm{eff}$)}, and the position of the
 peaks is shifted to smaller 
  radii. Inspecting the \tr{``dynamical phase diagram''} (\tb{point (e) vs. point
    (b)}), one finds that in case of point (e) much 
  more modes are unstable and grow considerably faster.
  \tr{Additionally, the value
  $1/\tau_\mathrm{m}(\capL)=\underset{k}{\max}(1/\tau(k,\capL))$ 
  of the maximum of $1/\tau$ (see
  Eq. (\ref{eq:taum})) almost reaches its upper limit
  $1=\underset{\capL}{\max}(1/\tau_{\rm m}(\capL))$.} 
  Both facts point towards a dynamics at large 
  scales, which is accelerated compared with the setup at higher
  temperatures. However, the 
  situation is different for $\kappa=40.0$. Here for point (f), the overall
  inbound motion is 
  slowed down even more \tb{than for point (c). (Fig.~\ref{fig:radial1}(c)
    corresponds to the same $\kappa$ but to a higher $T_\mathrm{eff}$.)}
   Again this may be explained on the basis of the
  \tr{outside panels} (e) and (f) for \tr{$1/\tau(k)$} in the ``dynamical phase
  diagram''. In both 
  cases, many modes are unstable, however, for point (f) the \tr{position of
    the} maximum 
  $1/\tau_\mathrm{m}$ of $1/\tau(k)$ 
  is shifted to large values of $k$, such that the nucleation of
  many small 
  clusters with a narrow size distribution dominates the dynamics. The modes
  corresponding to large scales also grow, but less pronounced relative to
  the fastest growing modes at larger $k$. In this low temperature limit, the
  dynamics realizes a ``bottom-up'' scenario for the evolution of  the
  system. The 
  nucleation of small clusters happens at much shorter time scales compared to
  the overall collapse. \tb{This is in} contrast to the ``top--down''
  scenario, anticipated for 
  small values of $\kappa$ or in the Newtonian limit. In this latter case\tr{,}
  \tb{particularly} at higher 
  temperatures, the formation of small clusters is unfavorable, whereas the
  overall growth of modes corresponding to the large scales sweeps  all
  particles to the final close packed patch. Inspection of panel (d) in
    Fig.~\ref{fig:radial1} finally reveals the onset of the formation of
    a shock wave at low temperature and at late times. The corresponding point
    in the phase diagram (point (d)) lies well below the line F. Recalling
    that systems with identical fastest growing modes are located on
    the same thin red line in Fig.~\ref{fig:phase}, we conclude that the
    fastest growing mode of po\tr{i}nt (d) is comparable to point
    (b)). However, the 
    value of  
    $\kappa=0.67$ is too small for the formation of a clearly visible shock
    wave. For that purpose, according to \tr{E}q. (\ref{eq:sw_pos}) the
    formation point of the shock wave has to lie at somewhat larger values of
    $\capL$, 
    corresponding to a radial extent beyond the size of the close packed final
    cluster. This is the case for interaction ranges corresponding to $\kappa
    \gtrsim 1.0$.  

 The picture is completed by one more setup \tb{at the boundary 
  between the gravitation--like, cold collapse and the transition regime
  marked by shock waves and in the vicinity of the line of stability
    (line S). This
  corresponds to point (g) in Fig.~\ref{fig:phase}.}  The radial
  evolution is shown in \tg{panel (g) of }Fig.~\ref{fig:radial1}. \tg{It}  
  \tg{corresponds to} point (g) \tg{and} shows the
  build--up and the evolution of the shock wave for $\kappa=4.0$ at a rather
  high temperature ($N=121$, $K^{-1}=0.097$, \tb{$T_\mathrm{eff}=0.52$}). The
  peak is almost ``molten'' in 
  the sense, that it becomes rather broad whereas the inbound motion of the
  peak is again accelerated. Concerning the \tb{location in the ``dynamical
    phase diagram'', point (g)} 
  is  
  \tg{located well above the line $\tau_{\rm f}=1.45$}
  and above
  the line F. It corresponds to a collective collapse at a high effective
  temperature. The overall inbound collective motion dominates the
  dynamics, whereas the high temperature disfavors the  
  formation of smaller clusters. \tg{The evolution is
    slowed down significantly.} \tb{(This is in contrast to point 
    (c) for the same value of $T_\mathrm{eff}$ but
  deeper in the spinodal regime (see Fig.~\ref{fig:radial1}(c)).)} Note that
\tb{point (g)} 
  is slightly outside the transition region where the
  shock wave is the dominant structure. The phenomenology is somewhat similar
  \tb{to the one for point (d) and the corresponding density evolution shown in
    panel \tg{(d)} of Fig.~\ref{fig:radial1}} ($\kappa=0.67$, $N=3844$,
  $K^{-1}=0.003$, \tb{$T_\mathrm{eff}=3.6\times 10^{-4}$}). Here, \korr{with}
  the onset of 
  the \tr{shock wave} \korr{becoming} visible only \korr{at}  
  late times, the dynamics corresponds to an almost collective collapse. We
  note\tr{,} however, that in this case the corresponding data point \tb{(d)} is
  below the line F indicating that the transition region is not limited by a
  sharp borderline.   

\section{Experimental prospects for varying the capillary length}
\label{sect:lambda}
 
In actual experiments, a change \tr{of} \tb{the capillary length
  $\lambda=L/\kappa=\sqrt{\gamma/(g \Delta\rho)}$} can be achieved 
by varying the  surface tension $\gamma\propto \lambda^2$, e.g., by adding
surfactants. In order to keep nonetheless both \tb{Jeans' time
  $\mathcal{T}=\gamma/(\Gamma f^2 \varrho_0)$ } 
and \tb{the amplitude of the capillary potential $V_0=f^2/(2\pi\gamma)$}
(Eq.~(\ref{eq:2part})) 
constant, which allows the investigation of the isolated role of $\lambda$ as
a range parameter, 
$\mathcal{T}$ and $V_0$ may be rescaled suitably by changing the particle
radius $R$ and \tmg{the particle number density $\varrho_0$ of the monolayer}, 
respectively. \tb{We assume that the external force $f$ on 
the colloids is caused by their own weight, so that $f\propto R^3$.}
\korr{The} rescaling   
\begin{equation}
  \lambda^{\prime} = \xi \lambda,\quad R^{\prime}=\psi R, \quad
  \varrho_0^{\prime}=\phi \varrho_0 , 
\end{equation}
\korr{together with}{\renewcommand{\baselinestretch}{1.0}\normalsize\footnote{The mobility is usually given by
  $\Gamma=1/(6\pi\eta R)$ where $\eta$ denotes the dynamic viscosity. For
  half immersed particles at an air--water interface, we \tr{use the
    approximation} 
  $\Gamma=1/(3\pi\eta R)$ instead~\cite{Bleibel:2011}.}} $\Gamma^{-1} \propto
R$, \korr{leads to a rescaling of} $\mathcal{T}$ and $V_0$: 
\begin{equation}
  \label{eq:A2}
  \tr{\mathcal{T}^{\,\prime}}=\frac{\xi^2}{\psi^5 \phi}\mathcal{T},\quad
  V_0^{\prime}=\frac{\psi^6}{\xi^2} V_0. 
\end{equation}
The requirements \tr{$\mathcal{T}=\mathcal{T}^{\,\prime}$} and $V_0=V_0^{\prime}$
can be fulfilled by \korr{choosing}  
\begin{equation}
  \phi=\psi=\sqrt[3]{\xi}.
\end{equation}
Thus a change of $\lambda$ by a factor $\xi$, while keeping $\mathcal{T}$ and
$V_0$ constant, requires moderate changes $\psi$ and $\phi$ of the particle
radius $R$ and \tr{of} the density $\varrho_0$ of the system,
respectively. Within a limited range, this appears to be experimentally
accessible because
it is state of the art to prepare colloids with designed radii, covering even
\tr{several} orders of magnitudes. Any change in density depends on the size of
the colloids in the sense that close packing defines an upper limit $R^{-2}$
for $\varrho_0$. Alternatively, one may allow for a moderate change of
$\mathcal{T}\propto \sqrt[3]{\xi}$ by keeping the density constant
($\phi=1$) and by using $V_0^{\prime}=V_0$, i.e., $\psi^6=\xi^2$ (see
Eq.~(\ref{eq:A2})). By expressing the time variable in terms of the
corresponding  
characteristic time $\mathcal{T}$, the various situations with fixed $V_0$ are
easily comparable. For the simulations presented throughout this work, we
\tr{have} only kept $f^2/\gamma$ constant and \tr{have} ignored a possible
dependence of $f$ on 
$\lambda$~\cite{Oettel:2008}. 
Therefore we \korr{have} left both $R$ and $\varrho_0$ unchanged. 
\tmg{Further discussions on the values of the relevant scales for
  the instability (Jeans' time $\jeansT$ and Jeans' length $K^{-1}$) as a
  function of the system parameters applicable to various experimental
  situations can be found in Ref.~\cite{Dominguez:2010}.}

\section{Summary and conclusions}
\label{sect:sum}

By studying a colloidal monolayer at a fluid interface and \korr{characterized
  by} capillary attraction we have provided the theoretical foundation for a
\tr{``dynamical phase diagram''} of a \tmg{paradigmatic system} governed by
long--ranged attractive 
interactions. The relevant parameters in this \tr{``dynamical phase diagram''}  
are an effective temperature (\tmg{proportional to the ratio of the energy of
  thermal motion and the capillary energy}) and the range of the interaction
\tmg{relative to the characteristic size of the system}. The latter
parameter provides a smooth connection between infinitely--ranged
attractive interactions, 
which in the present case corresponds to 2D Newtonian gravity, and
``short--ranged'', van-der-Waals-like attraction.
In this diagram (Fig. \ref{fig:phase}) we find four regions of
interest which we have explored via Brownian dynamics simulations and
\tmg{numerical computations within} dynamical density functional theory. Above
a critical temperature 
particles do not aggregate in spite of the attraction. Below this temperature
we have identified three dynamical 
regimes for clustering \tb{which, however, are not sharply
  demarcated}. At interaction ranges \tmg{much larger than the system size
  (``small system'' limit)}, we have found collective
evolution, a global collapse associated with the fast growth of the
\tmg{spatially most extended} modes, \tmg{analogous to} 
gravitational collapse.
For interaction ranges \tmg{much smaller than the system size (``large
  system'' limit)}, 
the dynamics is dominated by \tmg{local clustering analogous to}
spinodal decomposition; the nucleation of small clusters is overlaid
by a rather slow global dynamics, with a large separation of time scales.
In between we have predicted theoretically and have observed in simulations a
transition region. \tmg{For a finite--sized, disk--shaped distribution
  \korr{the dynamics}  is
  characterized by}  
an inbound traveling shock wave.  
For a spatially homogeneous \tmg{distribution with periodic boundary
  conditions}, this transition region reveals itself in that 
\tmg{observables} (such as the total number
of clusters) exhibit the onset of deviations in their temporal evolution from
that valid for systems with infinitely--ranged interactions~\cite{Bleibel:2011}.

\begin{appendix}

\section{Cold collapse solution and perturbation theory}
\label{app:pert}

\subsection{Lagrangian coordinates}
\label{sec:pert}

Here we recall briefly the Lagrangian formalism \tr{\texttt{(see, e.g.,
  Ref.~\cite{Buch95a})}} and introduce the notation to be used
throughout the Appendix. The key ingredient of the Lagrangian
formalism is the Lagrangian trajectory field of volume elements,
$\br_L (\bx,t)$, defined physically as the position $\br$ at time $t$
of that volume element which was at position $\bx$ initially (i.e., at
time $t = 0$); $\bx$ is called the Lagrangian coordinate of the volume
element.  This field defines a \tr{map}
\begin{equation}
  \label{eq:mapLagr}
  \bx \longmapsto \br = \br_L(\bx,t)
\end{equation}
which reduces to the identity at the initial time:
\begin{equation}
  \label{eq:initx}
  \br_L(\bx,0)=\bx .
\end{equation}
This \tr{map} is assumed to be invertible, i.e., the trajectories of
different volume elements do not cross each other{\renewcommand{\baselinestretch}{1.0}\normalsize\footnote{If this
  would happen, the corresponding system of equations would cease to
  have a unique solution. In this case, physical arguments would be
  required in order to extend the time interval of validity of the
  equations.}}.
The trajectory field allows one to express any other field as function
of the Lagrangian coordinates $\bx$ and to introduce the Lagrangian
fields for the density $(n_L)$, the long--ranged capillary potential $(w_L)$, the
reduced pressure $(\Pi_L)$, and the corresponding chemical potential ($\mu_L$):
\begin{subequations}
\label{eq:lagrfields}
\begin{equation}
  n_L(\bx, t) := n( \br=\br_L(\bx, t), t) ,
\end{equation}
\begin{equation}
  w_L(\bx, t) := w( \br=\br_L(\bx, t), t) ,
\end{equation}
\begin{equation}
  \Pi_L(\bx, t) := \Pi( \korr{n = n_L(\bx, t)} ) ,
\end{equation}
\begin{equation}
  \mu_L(\bx, t) := \mu( \korr{n = n_L(\bx, t)} ) .
\end{equation}
\end{subequations}
(We shall systematically use the subindex $L$ in order to indicate
that the corresponding field must be understood as a Lagrangian field,
i.e., as a function of $\bx$ and $t$, whenever the same field has been
already defined in Eulerian coordinates \korr{$\br$}.) In general, Lagrangian and
Eulerian fields will only coincide at the initial time.
As explained below, the equations for the density evolution (Eq.~(\ref{eq:euler})) 
can be transformed into the following equations
for the Lagrangian fields:
\begin{subequations}
  \label{eq:lagr}
\begin{equation}
  \label{eq:lagrM}
  \jacob (\bx, t) := ( \nabla_\bx \br_L)^{-1} ,
\end{equation}
\begin{equation}
  \label{eq:lagrx}
  \frac{d \br_L}{d t} = \bbg_L - (\jacob \cdot \nabla_\bx) \mu_L ,
\end{equation}
\begin{equation}
  \label{eq:lagrn}
  n_L = n_0 | \mathrm{det}\; \jacob | ,
\end{equation}
\begin{equation}
  \label{eq:lagrg}
  \bbg_L := (\jacob\cdot\nabla_\bx) w_L , 
\end{equation}
\begin{equation}
  \label{eq:lagrw}
  (\jacob\cdot\nabla_\bx) \cdot \bbg_L = - n_L + \capL^2 w_L .
\end{equation}
\end{subequations}
This is a complete set of equations for the determination of the
trajectory field $\br_L(\bx,t)$ and all the other fields.
$\nabla_\bx$ denotes the nabla operator with respect to the Lagrangian
coordinate $\bx$, while we use the customary notation $d/dt$ to
indicate the Lagrangian time derivative, i.e., derivative \textit{at
  fixed} $\bx$. The relationship with the Eulerian time derivative is \tr{a
direct} consequence of the definitions (Eq.~\ref{eq:lagrfields}) of
the Lagrangian fields:
\begin{equation}
  \label{eq:lagrdt}
  \frac{d }{d t} = 
  \frac{\partial }{\partial t} + 
  \frac{d \br_L }{d t} \cdot \nabla .
\end{equation}
The dyadic \korr{product 
$\jacob^{-1}(\bx,t)$} is  
the Jacobian matrix of the \tr{map}
$\bx\mapsto\br$, i.e., $(\jacob^{-1})_{ij} = \partial r_j/\partial x_i$.
With this matrix, the transformation of the nabla operator $\nabla_\bx$
is given by
\begin{equation}
  \label{eq:lagrdx}
  \nabla = \jacob(\bx,t)\cdot\nabla_\bx .
\end{equation}
\korr{(The central dot $\cdot$ denotes a scalar product and, more
  generally, contraction of tensorial indices if the dyadic product
  $\mathcal{M}$ is involved.)} In this manner, Eq.~(\ref{eq:lagrw})
follows immediately from Eq.~(\ref{eq:w}) after introducing the
acceleration field $\bbg := \nabla w$ induced by capillary forces (see
Eq.~(\ref{eq:lagrg})).
One notices that Eq.~(\ref{eq:nwithmu}) describes the advection of
volume elements by the velocity field $\bbg - \nabla \mu$ (which is
proportional to the acceleration in the overdamped approximation (see
Eq.~(\ref{eq:drag})) \korr{with the proportionality constant being one
  in terms of the dimensionless variables}). Therefore, the dynamical
equation obeyed by the trajectory of a volume element is
\begin{equation}
  \label{eq:lagrv}
  \frac{d \br_L}{d t} = \bbg - \nabla \mu ,
\end{equation}
which \tr{turns into} Eq.~(\ref{eq:lagrx}) \tr{upon} transforming the nabla
operators according to Eq.~(\ref{eq:lagrdx}). Finally, since due to
mass conservation the number of particles in a volume element, which
is followed along its trajectory, is a dynamical invariant, one has
\begin{equation}
  \tr{n(\br = \br_L (\bx, t),t) d^2\br = n(\bx, 0) d^2 \bx ,}
\end{equation}
and Eq.~(\ref{eq:lagrn}) follows immediately because the Jacobian of
the \tr{map} $\bx\mapsto\br$ is
\begin{equation}
  \left|\left| \frac{\partial \br_L}{\partial \bx} \right|\right| 
  = \frac{1}{\left|\mathrm{det}\;\jacob\right|} .
\end{equation}
Although Eq.~(\ref{eq:lagrn}) is the formal solution of
Eq.~(\ref{eq:n}), it will be useful to express Eq.~(\ref{eq:n}) (or
Eq.~(\ref{eq:nwithmu})) in terms of Lagrangian fields:
\begin{eqnarray}
  \label{eq:nevol}
  \frac{d n_L}{d t} 
  & \stackrel{\textrm{Eqs.~(\ref{eq:n},\ref{eq:lagrdt})}}{=} &
  - n \nabla \cdot \left[ 
    \nabla w - \frac{1}{n} \nabla \Pi \right] \nonumber \\
  & \stackrel{\textrm{Eqs.~(\ref{eq:w},\ref{eq:gibbsduhem})}}{=} &
  n^2 - \capL^2 n w + n \nabla^2 \mu \nonumber \\
  & \stackrel{\textrm{Eqs.~(\ref{eq:lagrfields},\ref{eq:lagrdx})}}{=} &
  n_L^2 - \capL^2 n_L w_L 
  \tr{+ n_L \jacob\cdot\nabla_\bx\cdot [ \jacob\cdot\nabla_\bx \mu_L ]} .
\end{eqnarray}
According to Eq.~(\ref{eq:lagrn}), a singularity in the density field,
\korr{that is, a vanishing Jacobian ($\Leftrightarrow\mathrm{det}\;\jacob=\infty$) of
  the Lagrangian--to--Eulerian map $\bx\mapsto\br$, indicates that the
  map becomes multivalued. This shows up geometrically as the
  mutual crossing of the Lagrangian trajectories \cite{Buch95a}}.
This is prevented generically by the effect of the pressure, i.e., by
the term involving $\mu$. The ``cold limit'' (setting $\Pi=0$ or,
equivalently, \tr{setting} $\mu\to\mathrm{constant}$) is thus a
singular limit because it reduces the order of the differential
equation (see Eq.~(\ref{eq:euler}) or Eq.~(\ref{eq:nevol})). It
becomes a first--order partial differential equation and the Lagrangian
trajectories coincide with the characteristic curves of
Eq.~(\ref{eq:n}): they are defined in the four--dimensional space
$(r_1, r_2, t, n)$ (with $\br=(r_1,r_2)$) as solutions of the set of
ordinary differential equations \cite{CoHi89}
\begin{equation}
  \label{eq:charcurve}
  dt = \frac{dr_1}{\partial w/\partial r_1}
  = \frac{dr_2}{\partial w/\partial r_2}
  = - \frac{dn}{n \nabla^2 w} ,
\end{equation}
which \tr{precisely turns into} Eqs.~(\ref{eq:lagrv}, \ref{eq:nevol})
\korr{upon} using $t$ in order to parametrize the characteristic curves. The
crossing of characteristic curves shows up in the solution as a ``a
shock wave''.
Therefore, the ``cold limit'' is not a uniformly valid approximation.
Whenever any of these singularities appear, one should keep in mind that
they are regularized physically by the effect of pressure.

\subsection{Radially symmetric evolution}

\label{sec:radsym}

\tr{Equation}~(\ref{eq:euler}) with the initial condition given by
Eq.~(\ref{eq:n0}) describe\tr{s} the collapse of the initial top--hat
profile under the combined action of the capillary attraction and the
gas pressure. 
The evolution preserves the radial symmetry of the initial
configuration, so that the Lagrangian trajectories are radial and must
have the form ($x := |\bx|$)
\begin{equation}
  \label{eq:radialx}
  \br_L(\bx, t) = a(x, t)\, \bx ,   
\end{equation}
with a certain amplitude function $a(x, t)$. Geometrically, this
describes the evolution of infinitesimally thin rings of matter with a
radius $a x$. In this case one has \korr{the dyadic product}
($\mathcal{I}$ is the identity 2nd--rank tensor)
\begin{equation}
  \label{eq:radialM}
  \jacob^{-1} = \nabla_\bx\br_L = a \mathcal{I} + \frac{\bx\bx}{x} \frac{\partial a}{\partial x} 
  = \left(
    \begin{array}[c]{cc}
      a + x (\partial a/\partial x) & 0 \\
      0 & a 
    \end{array}
  \right) ,
\end{equation}
where the last expression is the matrix representation \tr{of the dyadic}
in the basis $\{\be_r = \bx/x, \be_\varphi\}$ \korr{formed by} the unit vectors for
polar coordinates. 
Therefore, from Eq.~(\ref{eq:lagrn}) one obtains
\begin{equation}
  \label{eq:radialn}
  a^2 \left( 1 + \frac{x}{a} \frac{\partial a}{\partial x} \right) = 
  (\mathrm{det}\;\jacob)^{-1} = \frac{n_0(x)}{n_L(x,t)} ,
\end{equation}
with $n_0(x)$ given by Eq.~(\ref{eq:n0})\tr{. (At} the initial time one can
use $\br$ and $\bx$ interchangeably, see Eq.~(\ref{eq:initx})\tr{.)} 
This equation can be \tr{integrated rendering} 
\begin{equation}
  a(x,t) = \sqrt{\frac{A(t)}{x^2} + \frac{2}{x^2} \int_0^{x} d\xi \; \xi 
    \frac{n_0(\xi)}{n_L(\xi,t)}} ,
\end{equation}
where the integration constant $A(t)$ controls the \korr{shape} of the
trajectories near the origin \tr{so that}
\begin{equation}
  \label{eq:smallx0}
  \br_L (\bx\to 0, t) = a(x\to 0, t) \bx 
  = \frac{\bx}{x} \sqrt{A(t)} ,
\end{equation}
because the density field $n_L(\xi,t)$ must be smooth at $\xi=0$
before any singularity arises. Since the origin remains fixed during
the collapse, $\lim_{x\to 0} \br_L = {\bf 0}$, one must take
$A(t)\equiv 0$  
and
\begin{equation}
  \label{eq:radiala}
  a(x,t) = \sqrt{\frac{2}{x^2} \int_0^{x} d\xi \; \xi 
    \frac{n_0(\xi)}{n_L(\xi,t)}} .
\end{equation}

\subsubsection{Newtonian, cold limit}

\label{sec:radsym_cold}

If both $\capL$ and $\Pi$ vanish, the evolution reduces to the
collapse of the disk under its own ``gravitational'' attraction, for
which an analytically exact solution has been found (see, e.g.,
Refs.~\cite{Dominguez:2010,ChSi11}). In this case Eq.~(\ref{eq:nevol})
\korr{reduces} to
\begin{equation}
  \label{eq:Newtcoldn}
  \frac{d n_L}{d t} = n_L^2
  \quad\Rightarrow\quad
  n_L(\bx,t) = \frac{n_0(\bx)}{1-t n_0(\bx)}
\end{equation}
so that for a radially symmetric profile Eq.~(\ref{eq:radiala}) results in
\begin{equation}
  a(x, t) = \sqrt{ 1 - \frac{2 t }{x^2} \int_0^{x} d\xi \; \xi 
    \, n_0(\xi) } .
\end{equation}
(See also App.~C in Ref.~\cite{Dominguez:2010} for the radially
symmetric evolution in the Newtonian, cold limit.) 
For the initial condition in Eq.~(\ref{eq:n0}) this gives
\begin{equation}
  \label{eq:hata}
  a(x, t) =  \left\{
    \begin{array}[c]{cl}
      \displaystyle \sqrt{ 1 - t } , & x \leq 1 \\ 
      & \\
      \displaystyle \sqrt{ 1 - \frac{t}{x^2} } , & 1 < x 
    \end{array}
    \right.
\end{equation}
and
\begin{equation}
  \label{eq:hatn}
  n_L(\bx, t)  = \left\{
    \begin{array}[c]{cl}
      \displaystyle \frac{1}{1-t} , & x < 1 \\ 
      & \\
      0 , & 1 < x . 
    \end{array} \right.
\end{equation}
\korr{We note} that the Lagrangian trajectories are defined even in the vacuum
region ($x>1$). There they describe the trajectories of \textrm{test
  particles} in the sense of field theory, i.e., of a dilute particle
distribution which reacts to the force field without perturbing it.
One can also compute the ``gravitational'' field $w_L(\bx,t)$ as
the solution of Eq.~(\ref{eq:lagrw}), which \korr{reduces} to $\nabla_\bx^2
w_L = a^2(x,t) n_L$ and has the solution
\begin{equation}
  \label{eq:hatw}
  w_L(\bx,t) = C(t) - \frac{1}{4} x^2 
  \qquad (x \leq 1) ,
\end{equation}
where $C(t)$ is a, possibly time--dependent, additive constant.
(The value of the field in the vacuum region $x>1$ will not be
relevant for our purposes.)
Finally, the Eulerian density field can be obtained straightforwardly
from Eq.~(\ref{eq:lagrfields}):
\begin{equation}
  \label{eq:hatnEuler}
  n(\br,t) = \left\{
    \begin{array}[c]{cl}
      \displaystyle \frac{1}{1-t} , & |\br| < \sqrt{1-t} \\
      & \\
      0 , & \sqrt{1-t} < |\br| . \\
    \end{array} \right.
\end{equation}

Therefore, the top--hat profile collapses without deformation and
reaches an infinite density at time $t=1$, when all rings of matter
reach \textit{simultaneously} the center. (This is an example in which
the singularity is regularized by the gas pressure; the proper
description of the later stages of the collapse must take into
account the term $\Pi$ in Eq.~(\ref{eq:n}).)
This homogeneous collapse is the counterpart to the cosmological
scenario of a homogeneously contracting universe; the simultaneous
collapse at the center would correspond to the \korr{``big crunch''}.

\subsubsection{Cold limit and $\kappa \neq 0$. Perturbative approach}

\label{sec:radsym_screened}

Here we first consider the cold limit, $\Pi\to 0$, and investigate the
effect of a non-vanishing value of $\capL$. Equation~(\ref{eq:nevol}) is
now
\begin{equation}
  \label{eq:nevolcold}
  \frac{d n_L}{d t} = n_L^2 \left( 1 - \capL^2 \frac{w_L}{n_L} \right) ,  
\end{equation}
and cannot be solved exactly because of the 
dependence of $w_L$ on $n_L$. However, one can study the solution
perturbatively in the two opposite limits of ``small systems''
  ($\capL\ll 1$) and ``large systems'' ($\capL\gg 1$). The idea is to
  solve the approximate equation
\begin{equation}
  \label{eq:nevolcoldapprox}
  \frac{d n_L}{d t} \approx b(x, t) n_L^2 ,
  \quad
  b(x,t) := 1 - \capL^2 \frac{\hat{w}_L(x,t)}{\hat{n}_L(x,t)} ,
\end{equation}
inside the disk, $|x|\leq 1$, where $\hat{w}_L$ \tr{and} $\hat{n}_L$ are the
fields evaluated in a reference solution, i.e., the
  unperturbed evolution, \tr{which} depends on the limit one is studying and
  will be specified below in each case.  The solution \tr{of} this
equation inside the disk for the initial condition in
Eq.~(\ref{eq:n0}) is
\begin{equation}
  \label{eq:coldn}
  n_L(\bx, t) = [1 - B(x,t)]^{-1} ,
  \quad
  B(x,t) := \int_0^t ds\; b(x,s) 
  \qquad
  (x < 1) .
\end{equation}
The evolution is completely determined if Eq.~(\ref{eq:radiala}) is
\korr{used} to calculate the perturbed radial trajectories:
\begin{equation}
  \label{eq:colda}
  a(x, t) = \sqrt{ 1 - \frac{2}{x^2} \int_0^{x} d\xi\; \xi\, B(\xi, t)  } 
  \qquad
  (x<1).
\end{equation}
The explicit dependence of both $n_L$ and $a$ on $x$ implies that,
unlike in the Newtonian limit, the initial top--hat profile will
deform during the evolution and an inhomogeneous density field will
develop inside the collapsing disk. In particular, a singularity in
the density field can emerge earlier than the global collapse at the
center. The time $t_\mathrm{s}(x)$ for the \textit{s}ingularity at
which the density of the matter ring, which started at $x$, diverges
follows from Eq.~(\ref{eq:coldn}) and is given \tr{implicitly by}
\begin{equation}
  \label{eq:tsing}
  1 - B(x, t_\mathrm{s}) = 0.
\end{equation}

\bigskip

\noindent
$\bullet$ \textit{Small systems}, $\capL\ll 1$. In this limit
the reference, unperturbed state is the Newtonian, cold limit computed
previously
with the fields $\hat{n}_L(\bx,t)$ \tr{and} $\hat{w}_L(\bx,t)$ given by
Eqs.~(\ref{eq:hatn}) and (\ref{eq:hatw}), respectively:
\begin{subequations}
  \label{eq:hatnwlargek}
  \begin{equation}
    \hat{n}_L(\bx, t) = \frac{1}{1-t} ,
    \qquad
    (x < 1),
  \end{equation}
  \begin{equation}
    \hat{w}_L(\bx, t) = C(t) - \frac{1}{4} x^2 
    \qquad
    (x < 1).
\end{equation}
\end{subequations} 
However, the value of $C(t)$, which
is irrelevant in the Newtonian limit, must be determined as a function
of $\capL$ because the original Eq.~(\ref{eq:w}) is not invariant
under a shift by a constant in the ``potential'' $w$. The
solution of Eq.~(\ref{eq:w}) with the boundary
conditions given in Eq.~(\ref{eq:bc}) and for the Eulerian top--hat
profile in Eq.~(\ref{eq:hatnEuler}) is given exactly as
\begin{equation}
  \tr{w(\br,t) = \frac{1}{2\pi} \int_{r'<\sqrt{1-t}} d^2\br' \;} 
  \tr{\frac{1}{1-t} K_0 (\capL |\br-\br'|), }
\end{equation} 
so that
\begin{equation}
  \tr{C(t) = w_L(\bx=0,t) = w(\br=0,t) 
  = \left. \frac{1 - z K_1(z)}{z^2} \right|_{z=\capL\sqrt{1-t}} .}
\end{equation}
Expanding Bessel's function for $\capL\to 0$ one finally obtains
\begin{equation}
  \label{eq:approxC}
  \tr{C(t)  
  = \frac{1}{4} \left( 1 - 2 \gamma_\mathrm{e} 
    - 2 \ln \frac{\capL \sqrt{1-t}}{2} \right) 
  + \mathcal{O}(\capL^2) }
\end{equation}
($\gamma_\mathrm{e} = 0.5772\dots$ is the Euler--Mascheroni constant).
With this result, the density evolves according to
Eqs.~(\ref{eq:coldn}, \ref{eq:colda}) \tr{in terms of} the function 
\begin{equation}
  \label{eq:Bcorr}
  \tr{B(x,t) = t - \capL^2 \int_0^t ds \; (1-s) }
  \tr{\left[ C(s) - \frac{1}{4} x^2 \right] }
  \qquad (x < 1) 
\end{equation}
\tr{which is} obtained by inserting Eqs.~(\ref{eq:hatnwlargek}, \ref{eq:approxC})
into the definitions in
Eqs.~(\ref{eq:coldn}, \ref{eq:nevolcoldapprox}). We note that the ratio
$\hat{w}_L/\hat{n}_L$ \tr{vanishes} in time \tr{for} $t\to 1$, so that the
approximation leading to Eq.~(\ref{eq:nevolcoldapprox}) is uniformly
valid in time: if the term $\capL^2 \hat{w}_L/\hat{n}_L$ is a
perturbation at the initial time, it will \tr{consistently} remain so up to
the collapse at time $t=1$.

As \tr{stated} before, the emergence of a singularity is a generic outcome of
the model equation~(\ref{eq:nevolcold}). The time $t_s$ given
implicitly by Eq.~(\ref{eq:tsing}) approaches \tr{1} 
as $\capL \to 0$ (\tr{corresponding to} the simultaneous collapse of all
rings at the center), so that one can compute the time for the
occurrence of the singularity consistently within the perturbation
theory as
\begin{equation}
  t_\mathrm{s}(x) \approx  
  1 + \frac{\capL^2}{8} \left[ \frac{3}{2} - 2 \gamma_\mathrm{e} - \ln\frac{\capL^2}{4} - x^2 \right] ,
\end{equation}
after evaluating the integral in Eq.~(\ref{eq:Bcorr}) at $t=1$. The
physically meaningful time is the earliest one which marks the
appearance of the first shock wave and the limit of validity of the
cold approximation. This earliest time corresponds to $x=1$, i.e., the
outer rim of the disk, so that the time of emergence of the \textit{s}hock
\textit{w}ave occurs at 
\begin{equation}
  \label{eq:tswsmallkappa}
  t_\mathrm{sw} = t_\mathrm{s}(x=1) \approx 
  1 + \frac{\capL^2}{8} \left[ \frac{1}{2} - 2 \gamma_\mathrm{e} - \ln\frac{\capL^2}{4} \right] .
\end{equation}
The position of appearance of \tr{this} shock wave is given by the outer rim
of the disk\tr{:}
\begin{equation}
  \label{eq:rswsmallkappa}
  r_\mathrm{sw} = |\br_L (x=1, t=t_\mathrm{sw})| = a (1, t_\mathrm{sw}) 
  \approx \frac{\capL}{4} , 
\end{equation}
to lowest order in $\capL$ according to Eq.~(\ref{eq:colda}).
Figure~\ref{fig:theorySW}\tr{(a)} shows an example of the density profile
\tr{evolving} according to \tr{this} perturbative calculation.

\bigskip

\noindent
$\bullet$ \textit{Large systems}, $\capL\gg 1$. In this limit
  the reference state, which enters into
  Eq.~(\ref{eq:nevolcoldapprox}) via $b(x,t)$, is the initial
  configuration, 
  \begin{subequations}
    \label{eq:hatnwlargekappa}
    \begin{equation}
      \tr{\hat{n}_L(x,t) = n_0(x) = 1, 
      \qquad
      (x<1) ,}
    \end{equation}
    \begin{equation}
    \tr{\hat{w}_L(x,t) = w_L(x,t=0) = C(0) - \frac{1}{4}x^2
    \qquad
    (x<1) ,}
  \end{equation}
  \end{subequations}
  because $d n_L/d t$ \korr{can be neglected} due to Eqs.~(\ref{eq:largekappa}),
  \tr{(\ref{eq:nevolcold}), and (\ref{eq:nevolcoldapprox})}. This simply
  indicates that the evolution proceeds over a time scale much larger
  than Jeans' time. The only exception is the behavior near the rim,
  where the discontinuity in the initial density renders the
  approximation given by Eq.~(\ref{eq:largekappa}) invalid. To be more
  specific, one can solve Eq.~(\ref{eq:w}) for the initial condition
  for $n$ \korr{as given by Eq.~(\ref{eq:n0})}:
  \begin{equation}
    \label{eq:initw}
    \capL^2 w (r, t=0) = \left\{
      \begin{array}[c]{cl}
        1 - \capL K_1 (\capL) I_0(\capL r) , & r < 1 \\
        & \\
        \capL I_1(\capL) K_0(\capL r) , & 1 \leq r .
      \end{array}
    \right.
  \end{equation}
  Since Lagrangian and Eulerian coordinates coincide at the initial
  time (see Eq.~(\ref{eq:initx})), one can use this result in
  Eq.~(\ref{eq:nevolcoldapprox}) for $b(x,t)$ and obtains, with the
  additional approximation of large $\capL$,
\begin{equation}
  \label{eq:approxb}
  b(x,t) \approx \left\{
    \begin{array}[c]{cl}
      \displaystyle \frac{1}{2} \mathrm{e}^{-\capL (1-x)} , & x < 1 \\
      & \\
      \displaystyle \frac{1}{2} + \frac{1}{4\capL} , & x = 1^{-}
    \end{array}
  \right.
  \qquad
  (\capL \gg 1) 
\end{equation}
inside the disk, so that $B(x,t) = t b(x,t)$ in the perturbative
solutions (Eqs.~(\ref{eq:coldn}, \ref{eq:colda})).
As in the previous case, a \textit{s}ingularity appears at a time
$t_\mathrm{s}$ given by Eq.~(\ref{eq:tsing}); in this case
\begin{equation}
  \label{eq:tsinglargekappa}
  t_\mathrm{s}(x) = \left\{
    \begin{array}[c]{cl}
      \displaystyle 2 \mathrm{e}^{\capL (1-x)}, & x < 1 \\
      & \\
      \displaystyle 2 - \frac{1}{\capL} , & x=1^-
    \end{array}
  \right.
\end{equation}
indicating the occurrence of two widely separated time scales in the
evolution: one very slow compared with Jeans' time strictly inside the
disk, $x<1$, and a fast one at the rim, $x=1$, where the shock wave,
i.e., the earliest singularity, forms at a time $t_\mathrm{sw} \approx
2-1/\capL$ (compare with Eq.~(\ref{eq:tswsmallkappa})). Nevertheless,
in both cases the radial displacement given by Eq.~(\ref{eq:colda}) is
very small,
\begin{equation}
  \label{eq:coldalargekappa}
  a(x,t) \approx 1 - \frac{t}{\capL t_\mathrm{s}(x)} ,
  \qquad
  \frac{t}{t_\mathrm{s}(x)} \leq 1 ,
\end{equation}
so that the physically meaningful singularity at the rim has a
position $r_\mathrm{sw} = a(1, t_\mathrm{sw}) \approx 1 - 1/\capL$.
Figure~\ref{fig:theorySW}\tr{(b)} illustrates the evolution of the density
profile according to \tr{this} perturbative calculation.
Formally speaking, \korr{for points with $x<1$} the approximation in
Eq.~(\ref{eq:approxb}) is a short--time expansion with respect to the
time scale of evolution (see Eq.~(\ref{eq:tsinglargekappa})).  Since
this scale is much larger than Jeans' time, the approximation is
\tr{also} valid for describing the evolution on Jeans' time scale.

\bigskip

In summary, a nonzero value of $\capL$ induces a deformation of the
initial top--hat profile by forming an enhanced density peak at the
outer rim. For any value of $\capL$ the time scale for forming the
singularity is of the order of Jeans' time, but the \tr{radial position of
the peak} depends strongly on the value of $\capL$. The singularity
eventually \korr{evolves} into a shock wave when the effect of pressure
becomes relevant locally.

\subsubsection{``Hot'' effects ($\,\,\Pi\neq 0$)}
\label{sec:Newthot}

Here we add a few remarks concerning
the effect of the pressure $\Pi$ in Eq.~(\ref{eq:n}) on the radial
evolution. Generically, the cold limit $\Pi\to 0$ is singular 
because $\Pi$ is associated with the highest order of the spatial
derivatives in Eq.~(\ref{eq:n}), so that one cannot neglect the term
$\nabla\Pi$ uniformly throughout the disk. There are two particular
aspects fo which the effect of $\Pi$ is relevant, that is the jump of
the initial density at the outer rim of the disk and the formation of
the shock wave, because both features 
\korr{are associated with} the formal behavior $\nabla\Pi\to\infty$.

The initial discontinuity at $r=1$ (see Eq.~(\ref{eq:n0})) implies
that locally the early stages of the evolution will be dominated by
the regularizing effect of the pressure, no matter how small $\Pi$ is.
In order to be specific we consider the ideal gas approximation, $\Pi
= \varepsilon n$, in which the small parameter $\varepsilon\to 0$ can
be identified with a dimensionless temperature (see
Eq.~(\ref{eq:dimless})). The very early evolution of the density
discontinuity can be described by neglecting locally any other force
but pressure, so that in Eulerian coordinates
\begin{equation}
  \frac{\partial n}{\partial t} \approx \nabla^2 \Pi 
  = \varepsilon \frac{1}{r} \frac{\partial}{\partial r}
  \left( r \frac{\partial n}{\partial r} \right) ,
  \qquad
  r\approx 1 .
\end{equation}
In order to analyze the effect of pressure we introduce a new
independent variable $z := (r - 1)/\sqrt{\varepsilon}$ (centered at
the initial density jump and rescaled in proportion to the, as it will
turn out, thickness of the regularized discontinuity). This simplifies
the equation as $\varepsilon\to 0$ at finite values of $z$ (i.e., one
effectively neglects the curvature of the regularized discontinuity
and approximates it locally by a straight line):
\begin{equation}
  \frac{\partial n}{\partial t} \approx \frac{\partial^2 n}{\partial z^2} ,
  \quad
  n(z,t=0) = \left\{
    \begin{array}[c]{cl}
      1 , & z < 0 \\
      0 , & 0 < z .
    \end{array}\right.
\end{equation}
Expressed in terms of the original variables, the solution is given by
the error function:
\begin{equation}
  \label{eq:kink}
  n(r,t) \approx \frac{1}{2} \left[ 
    1 - \mathrm{erf}\left(\frac{r-1}{\sqrt{4\varepsilon t}}\right) 
  \right]
  = \frac{1}{2} \left[ 
    1 - \mathrm{erf}\left(\frac{z}{\sqrt{4t}}\right) 
  \right] .
\end{equation}
This represents a regularized jump discontinuity or kink of thickness
$\sim \sqrt{\varepsilon t}$. One can use this solution in order to
determine self--consistently \korr{the time $t_p$} beyond which the contribution
of the \textit{p}ressure term no longer dominates Eq.~(\ref{eq:n}) even
near $r=1$ compared \tr{with} the term $\nabla\cdot(n \nabla w)$, which is
initially of order unity \korr{in terms of dimensionless quantities}:
\begin{equation}
  \mathrm{max}_{r} \left| \varepsilon \frac{1}{r} \frac{\partial}{\partial r}
    \left( r \frac{\partial n}{\partial r} \right) \right| 
  \approx \mathrm{max}_{z} \left| \frac{\partial^2 n}{\partial z^2} \right| \lesssim 1 
  \quad\Rightarrow\quad
  t_p \gtrsim (8\pi\mathrm{e})^{-1/2} \approx 0.12 \tr{.}
\end{equation}
The fact that this time is independent of $\varepsilon$ means that the
regularization of the initial discontinuity takes a small but finite
fraction of the total time of collapse in the Newtonian limit, which
is \tr{of} order 1 (see Subsec.~\ref{sec:radsym_cold} and
Fig.~\ref{fig:theorySW}). However, this effect is  
spatially localized because the thickness of the resulting
kink (Eq.~\ref{eq:kink}) vanishes as $\varepsilon\to 0$.

In the same manner, when a shock wave appears 
the density gradients can become so large that the effect of a small
pressure eventually dominates the evolution in the localized regions
of large gradients. This is the hallmark of the formation of a
\emph{moving boundary layer}. The involved mathematical analysis of
this phenomenon is beyond the scope of the present study; here we
confine our effort to the insight
that this mechanism provides a qualitative explanation of the
regularization of shock--waves as observed in simulations.

\subsection{\tr{Non-radially symmetric perturbations}}
\label{sec:nonrad}

We now explore the robustness of the radially symmetric evolution
against small perturbations of the initial radial symmetry. In this
respect the ultimate goal is a generalization of the stability
analysis as already \tr{carried out} for the static \tr{(i.e.,
  time--independent)} homogeneous background (see \korr{Subsec.}~\ref{sec:hom}).
However, the mathematical problem is substantially more complicated,
given that the unperturbed state, i.e., the radially symmetric density
profile studied in the previous \tr{s}ections, is neither static nor
homogeneous.
Lagrangian perturbation theory has proved to be successful in the
study of cosmological structure formation in an expanding homogeneous
universe \cite{Buch95a,AdBu99} and, as it will turn out below, it is
also helpful for the present problem. But, compared to 
applications \korr{in cosmology}, progress is limited due to two
difficulties: (i) the presence of the initial inhomogeneity at the
boundary of the disk, and (ii) corrections to the Newtonian limit.
Therefore, here we \tr{can address}
the full stability analysis \tr{only partially.}

\subsubsection{Newtonian, cold limit}

\tr{We first} consider the exact solution in Eq.~(\ref{eq:Newtcoldn})
specialized to an initial density distribution of the form
\begin{equation}
  n_0 (\bx) = \left\{
    \begin{array}[c]{cl}
      1 + \delta n_0 (\bx)  , & x < 1 \\
      & \\
      \delta n_0 (\bx) , & 1 < x , \\
    \end{array} \right.
\end{equation}
i.e., a perturbation of
the top--hat profile given by Eq.~(\ref{eq:n0}).
One can distinguish two cases:
\begin{enumerate}
\item For perturbations inside the disk ($x<1$) one has
  \begin{equation}
    \label{eq:frag}
    n_L(\bx,t) = \frac{1 + \delta n_0(\bx)}{1-t [ 1 + \delta n_0(\bx) ] } ,
  \end{equation}
  and a density singularity arises at the time $t_\mathrm{s}(\bx) =
  [1 + \delta n_0(\bx)]^{-1}$. Therefore any local overdensity
  ($\delta n_0 > 0$) will grow faster and collapse earlier than the
  disk as a whole.  This is the counterpart of the so--called
  \textit{``fragmentation instability''} of a collapsing spherical
  cloud in astrophysics \cite{Hunt62,ALP88}.
\item For perturbations outside the disk ($x>1$), representing in
  particular \korr{deviations} from the circular shape of the initial disk,
  one has
  \begin{equation}
    n_L(\bx,t) = \frac{\delta n_0(\bx)}{1-t \delta n_0(\bx) } ,
  \end{equation}
  and a density singularity arises at \tr{the} time $t_\mathrm{s}(\bx) =
  1/\delta n_0(\bx)$. For small perturbations ($0<\delta n_0 \ll 1$),
  this time is larger than the time of collapse of the disk.
\end{enumerate}
We emphasize that these results are exact, which is a peculiarity of
the overdamped dynamics. In the astrophysical counterpart of this
problem, the relevance of inertia prevents one from finding a closed
equation for the density field analogous to Eq.~(\ref{eq:Newtcoldn})
and thus it is unavoidable to resort to perturbation theory.
Nevertheless, the problem is not fully solved until the
Lagrangian-to-Eulerian \tr{map} $\br_L (\bx,t)$ has been computed.
Since this cannot be done exactly it is useful to apply Lagrangian
perturbation theory.
To this end, one defines a perturbation 
\begin{equation}
  \delta \br_L(\br, t) := \br_L(\br, t) - \hat{\br}_L(\bx,t) ,
\end{equation}
of the trajectory, where the \tr{hat}  
over a symbol 
denotes the reference, unperturbed evolution. The initial condition
\begin{equation}
  \label{eq:initpertx}
  \delta \br_L (\bx, t=0) = {\bf 0} 
\end{equation}
holds by definition of the Lagrangian \tr{map} (see
Eq.~(\ref{eq:initx})).
Likewise, one defines perturbations of the Lagrangian fields as
\begin{equation}
  \delta n_L (\bx, t) := n_L (\bx, t) - \hat{n}_L (\bx, t) ,
\end{equation}
and similarly for $\delta \jacob$, $\delta\bbg_L$, $\delta w_L$, and
$\delta\mu_L$. \tr{Equation}~(\ref{eq:lagr}) can be linearized with
respect to the small perturbations{\renewcommand{\baselinestretch}{1.0}\normalsize\footnote{Linearization of terms
  involving the matrix $\jacob$ is achieved by applying the identities
  $\ln \mathrm{det}\; \jacob = \mathrm{tr}\;\ln \jacob$ and
  $\jacob^{-1} = \sum_{n=0}^{\infty} (\mathcal{I} - \jacob)^n$, where
  $\mathcal{I}$ is the identity matrix \korr{and $\mathcal{M}^{-1} =
    \nabla_\bx \br_L$ (Eq.~(\ref{eq:lagrM}))}.}}:
\begin{subequations}
  \label{eq:pert}
  \begin{equation}
    \delta \jacob \approx - \hat{\jacob}\cdot (\nabla_\bx \delta\br_L ) \cdot \hat{\jacob} ,
  \end{equation}
  \begin{equation}
    \label{eq:pertx}
    \frac{d \delta \br_L}{d t} \approx \delta \bbg_L 
    - (\hat{\jacob} \cdot \nabla_\bx) \delta \mu_L 
    - (\delta \jacob \cdot \nabla_\bx) \hat{\mu}_L ,
  \end{equation}
  \begin{equation}
    \label{eq:pertn}
    \delta n_L \approx | \mathrm{det}\; \hat{\jacob} | 
    \left[
      \delta n_0 + \hat{n}_0 \mathrm{tr}\; (\hat{\jacob}^{-1} \cdot \delta\jacob )
    \right] ,
  \end{equation}
  \begin{equation}
    \label{eq:pertmu}
    \delta \mu_L \approx \hat{\mu}'_L \delta n_L
    = \frac{\hat{\Pi}'_L}{\hat{n}_L} \delta n_L ,
  \end{equation}
  \begin{equation}
    \label{eq:pertg}
    \delta \bbg_L \approx (\hat{\jacob}\cdot\nabla_\bx) \delta w_L 
    + (\delta\jacob\cdot\nabla_\bx) \hat{w}_L , 
  \end{equation}
  \begin{equation}
    \label{eq:pertw}
    (\hat{\jacob}\cdot\nabla_\bx) \cdot \delta \bbg_L 
    + (\delta\jacob\cdot\nabla_\bx) \cdot \hat{\bbg}_L \approx - \delta n_L + \capL^2 \delta w_L ,
  \end{equation}
\end{subequations}
where $\hat{\mu}'_L := d\mu/d \hat{n}_L$, $\hat{\Pi}'_L := d\Pi/d
\hat{n}_L$, $\delta n_0(\bx) := \delta n_L(\bx, 0)$ represents the
initial density perturbation, and $\hat{n}_0(\bx) := \hat{n}_L(\bx,0)$
is the initial unperturbed density.
These equations simplify significantly when specialized to the
perturbed evolution of a top--hat profile in the Newtonian, cold
limit, which amounts to setting $\capL\to 0$, $\mu_L\to 0$, and,
restricting attention to the interior of the disk ($x<1$),
\begin{equation}
  \label{eq:refTopHat}
  \hat{n}_L = \tr{\frac{1}{\hat{a}^2} ,}
  \quad
  \hat{\mathcal{M}} = \tr{\frac{1}{\hat{a}}} \mathcal{I},
  \quad
  \tr{\hat{w}_L = C(t) - \frac{1}{4} x^2 ,}
\end{equation}
\tr{with $\hat{a}(t) := \sqrt{1-t}$ (see
Eqs.~(\ref{eq:radialM}, \ref{eq:hatn}, \ref{eq:hatw})).}
Under these conditions one obtains
\begin{subequations}
  \begin{equation}
    \delta \jacob \approx - \frac{1}{\hat{a}^2} \nabla_\bx \delta\br_L ,
  \end{equation}
  \begin{equation}
    \label{eq:pertNewtr}
    \frac{d \delta \br_L}{d t} \approx \delta \bbg_L ,
  \end{equation}
  \begin{equation}
    \label{eq:pertNewtn}
    \delta n_L \approx \frac{1}{\hat{a}^2} 
    \left[
      \delta n_0 - \frac{1}{\hat{a}} \nabla_\bx \cdot \delta\br_L 
    \right] ,
  \end{equation}
  \begin{equation}
    \label{eq:pertNewtg}
    \delta \bbg_L \approx \frac{1}{\hat{a}} \nabla_\bx \delta w_L 
    + \frac{1}{2\hat{a}^2} (\nabla_\bx \delta\br_L) \cdot \bx ,
  \end{equation}
  \begin{equation}
    \frac{1}{\hat{a}} \nabla_\bx \cdot \delta \bbg_L 
    + \frac{1}{2\hat{a}^3} (\nabla_\bx \cdot \delta\br_L)
    \approx - \delta n_L .
  \end{equation}
\end{subequations}
In order to proceed one introduces the auxiliary field
\begin{equation}
  \label{eq:phi}
 \phi_L := \delta w_L + \frac{1}{2\hat{a}} \delta\br_L \cdot \bx ,
\end{equation}
which can be viewed as a potential in Lagrangian coordinates:
$\nabla_\bx \phi_L = \hat{a} \delta \bbg_L + \delta\br_L / \tr{(2\hat{a})}$.
Equations~(\ref{eq:pertNewtr}, \ref{eq:pertNewtg}) reduce to
\begin{subequations}
  \label{eq:pertNewt}
  \begin{equation}
    \frac{d \delta \br_L}{d t} \approx \frac{1}{\hat{a}} \nabla_\bx \phi_L 
    - \frac{1}{2 \hat{a}^2} \delta\br_L 
  \end{equation}
  and
  \begin{equation}
    \nabla_\bx^2 \phi_L 
    \approx - \delta n_0 + \frac{1}{\hat{a}} \nabla_\bx \cdot \delta\br_L ,
  \end{equation}
\end{subequations}
from which one obtains{\renewcommand{\baselinestretch}{1.0}\normalsize\footnote{\korr{The vector field $\delta\br_L$
    and $\nabla_\bx$ consist of two components. Both can be extended
    to three--component vectors by adding a third (``vertical'')
    component taken to be zero. This way $\nabla_\bx\times\delta\br_L$
    is well defined and its only nonzero component is in the
    ``vertical'' direction. Accordingly, Eq.~(\ref{eq:rotdeltar})
    reduces to a single equation for the ``vertical'' component.}}}
\begin{subequations}
\label{eq:pertdeltar}
\begin{equation}
  \label{eq:rotdeltar}
  \frac{d}{d t} (\nabla_\bx \times\delta \br_L) \approx 
    - \frac{1}{2 \hat{a}^2} \nabla_\bx\times \delta\br_L ,
\end{equation}
and
\begin{equation}
  \frac{d}{d t} (\nabla_\bx \cdot\delta \br_L) \approx 
  - \frac{\delta n_0}{\hat{a}} 
  + \frac{1}{2 \hat{a}^2} \nabla_\bx\cdot\delta\br_L .
\end{equation}
\end{subequations} 
These equations can be integrated with the initial
condition given in Eq.~(\ref{eq:initpertx}):
\begin{subequations}
  \label{eq:pertNewtx}
\begin{equation}
  \label{eq:rotNewt}
  \nabla_\bx\times\delta\br_L (\bx, t) 
  \korr{= \hat{a}(t) \nabla_\bx\times\delta\br_L (\bx, t=0)}
  = \mathbf{0} ,
\end{equation}
\begin{equation}
  \nabla_\bx\cdot\delta\br_L (\bx, t) = 
  - \frac{1-\hat{a}^2(t)}{\hat{a}(t)} \delta n_0(\bx) .
\end{equation}
\end{subequations}
These equations indicate that the mass clusters at the initially
overdense regions ($\delta n_0 > 0$; note that $1-\hat{a}^2 \geq 0$)
while, consistently, the underdense regions ($\delta n_0 < 0$) get
depleted. Actually, the perturbation\tr{s} of the trajectories \tr{are} parallel
to the gravitational field \tr{generated} by the initial density
perturbation\tr{. This} follows from the form of Eq.~(\ref{eq:pertNewtx}):
the time dependence induced by $\hat{a}(t)$ can be factored out by a
simple rescaling of the perturbation $\delta\br_L$, and the resulting
equations are formally the field equations for the Newtonian
gravitational field \tr{generated} by a mass distribution given by $\delta
n_0(\bx)$. This property is the counterpart of the so--called
\textit{Zel'dovich approximation} in the context of cosmological
structure formation (see Ref.~\cite{SaCo95} and references therein).

\subsubsection{Small--scale, central perturbations}

The theoretical analysis beyond the cold, Newtonian limit is hampered
by the fact that even the evolution of the reference, unperturbed
state cannot be obtained in closed form. However, by focusing on
perturbations localized close to the center of the disk (``central
perturbations''), one can neglect the influence of the boundary so
that one can take advantage of the results presented in
\tr{Subsec.}~\ref{sec:hom}.

The simplest case is the limit of large system size, i.e., $\capL\gg
1$. As argued in \tr{Subsec.}~\ref{sec:radsym_screened}, in this case the time
scale of disk collapse is much larger than Jeans' time and, as far as
the central perturbations are concerned, one can approximate the disk
as a \tr{\textit{time--independent}} homogeneous distribution, namely the initial
\korr{one}, and the analysis presented in \tr{Subsec.}~\ref{sec:hom} holds.
Therefore, according to Fig.~\ref{fig:phase}, the fastest growing modes
are the ones on small spatial scales and they are characterized by
time scales of the order of Jeans' time (barring the exceptional case
$T_\mathrm{eff}\to 1$), which is consistent with the underlying
approximations.

In the opposite limit of small system size, i.e., $\capL\ll 1$, we can
establish a connection with the results presented recently in
Ref.~\cite{Chav11}, which deals, in the present language, with the
stability in the Newtonian limit ($\capL=0$) of an infinitely extended
disk (i.e., no boundary is considered).
In this case one can argue that, for the central perturbations, the
evolution of the reference state 
is indistinguishable from the cold, Newtonian limit during \tr{a certain}
initial period of time, because we have demonstrated before that the
effect of the disk boundary is localized and \tr{that} it takes some time \korr{for it} to
propagate into the interior of the disk. Thus, at early times
Eq.~(\ref{eq:pert}) can be evaluated in the region $x\ll 1$ \tr{by again using}
\tr{Eq.~(\ref{eq:refTopHat}):}
\begin{subequations}
  \begin{equation}
    \frac{d \delta \br_L}{d t} \approx \frac{1}{\hat{a}} \nabla_\bx \phi_L 
    - \frac{1}{2 \hat{a}^2} \delta\br_L 
    + \frac{\Pi' (\hat{n}_L)}{\hat{a}^2} \nabla_\bx (\nabla_\bx\cdot\delta\br_L ) ,
  \end{equation}
  \begin{equation}
    \label{eq:nablaphi2}
    \nabla_\bx^2 \phi_L 
    \approx - \delta n_0 + \frac{1}{\hat{a}} \nabla_\bx \cdot \delta\br_L 
    + \hat{a}^2 \capL^2 \left[ \phi_L - \frac{1}{2\hat{a}} \bx\cdot\delta\br_L \right] 
  \end{equation}
\end{subequations}
(compare with Eq.~(\ref{eq:pertNewt})). Furthermore, in order to be
consistent with this approximation one has to neglect the term
$\propto\capL^2$ in Eq.~(\ref{eq:nablaphi2}) because only
perturbations with length scales much shorter than $\capL^{-1}\gg 1$
are to be addressed. This leads to
\begin{subequations}
\begin{equation}
  \label{eq:pertrot}
  \tr{\frac{d}{d t} (\nabla_\bx \times\delta \br_L)} \approx 
    - \frac{1}{2 \hat{a}^2} \nabla_\bx\times \delta\br_L 
\end{equation}
and
\begin{equation}
  \label{eq:pertdiv}
  \tr{\frac{d}{d t} (\nabla_\bx \cdot\delta \br_L) \approx} 
  - \frac{\delta n_0}{\hat{a}} 
  + \frac{1}{2 \hat{a}^2} \nabla_\bx\cdot\delta\br_L 
  + \frac{\Pi' (\hat{n}_L)}{\hat{a}^2} \nabla_\bx^2 (\nabla_\bx\cdot\delta\br_L ) 
\end{equation}
\end{subequations}
(compare with Eq.~(\ref{eq:pertdeltar})). \tr{Integration of}
Eq.~(\ref{eq:pertrot}) implies
$\nabla_\bx\times\delta\br_L=\mathbf{0}$ \tr{due to} the initial condition
given in Eq.~(\ref{eq:initpertx}) \korr{(compare with Eq.~(\ref{eq:rotNewt}))}, 
while Eq.~(\ref{eq:pertdiv}) can be
written in a more familiar form by introducing the so--called
\textit{density contrast}, defined as
\begin{equation}
  \delta_L(\bx,t) := \frac{\delta n_L(\bx,t)}{\hat{n}_L(\bx,t)} 
  \approx \hat{a}^2 \delta n_L 
  = \delta n_0 - \frac{1}{\hat{a}} \nabla_\bx\cdot\delta\br_L ,
\end{equation}
so that
\begin{equation}
  \hat{a}^2 \frac{d \delta_L}{d t} \approx 
  \delta_L 
  + \Pi' (\hat{n}_L) 
  \nabla_\bx^2 \delta_L .
\end{equation}
By introducing the Fourier transform in Lagrangian coordinates,
\korr{$\tilde{\delta}_{L,\bk} := \int d^2\bx\; \delta_L(\bx) \textrm{e}^{-i\bk\cdot\bx}$},
one obtains
\begin{equation}
  \label{eq:centraldelta}
  \hat{a}^2 \frac{d \tilde{\delta}_{L,\bk}}{d t} \approx 
  \left[ 1 - \left(\frac{k}{K(t)}\right)^2 \right] \tilde{\delta}_{L,\bk}
\end{equation}
in terms of the time--dependent Jeans\tr{'} length
\begin{equation}
  \label{eq:timeK}
  K^{-1}(t) := \sqrt{\frac{d \Pi}{d n} (n= \hat{a}^{-2}(t))} .
\end{equation}
This is the same expression as the one obtained for the stability of
the homogenous state (see Eqs.~(\ref{eq:deltalin},
\ref{eq:tau})), 
restricted to modes with $k \gg 1 \gg \capL$ (i.e., valid only for
perturbations \korr{of} small length scales near the center of the disk).
The only difference is 
the explicit time dependence of $K$ introduced by the background
density $\hat{n}_L(t)$. For realistic equations of state, e.g., for
hard disks, Jeans' length will decrease in time
and in Fig.~\ref{fig:phase} the effective state of the system for
these ``central perturbations'' would describe a trajectory of
steadily increasing $T_\mathrm{eff} \propto K^{-2}(t)$. Thus, there is
an increasing number of modes at small scales for which the effect of
pressure will counteract the fragmentation
instability. 
The solutions of Eq.~(\ref{eq:centraldelta}) have been studied
recently by Chavanis \cite{Chav11} for the particular choice $\Pi(n)
\propto n^\alpha$ as function of the polytropic index $\alpha$. The
main conclusion follows from evaluating the
time--dependent Jeans\tr{'} length (see Eq.~(\ref{eq:timeK})), $K^2 (t)
\propto \hat{a}^{2(\alpha-1)}$, so that the amplitude of an initially
unstable mode ($k < K(t=0)$) will eventually die \korr{out} after a certain
time ($k > K(t)$) only if $\alpha > 1$ (corresponding to the critical
index $\gamma_{4/3}$ \tr{introduced in} Ref.~\cite{Chav11} for the spatial dimension
$d=2$){\renewcommand{\baselinestretch}{1.0}\normalsize\footnote{As pointed out in Ref.~\cite{Chav11}, this conclusion
  only holds provided the initial amplitude of the unstable mode is
  sufficiently small so that its growth and subsequent decay can be
  described by the linearized theory. If \korr{after a certain} time the amplitude
  enters the nonlinear regime, the conclusion concerning the
  asymptotic stability is no longer reliable.}}.

\end{appendix}

\begin{acknowledgments}
\label{acknowl}
J.B. thanks the German Research Foundation (DFG) for the financial
support through the Collaborative Research Center (SFB-TR6) ``Colloids
in External Fields'' Project No.~N01. \tmg{A.D. acknowledges support
  by the Spanish Government through Grants No.~AIB2010DE-00263 and
  No.~FIS2011-24460 (partially financed by FEDER funds).}

\end{acknowledgments}




\begin{thebibliography}{00}

\bibitem{Oettel:2008}
  M.~Oettel and S.~Dietrich,
  Langmuir \textbf{24}, 1425 (2008).

\bibitem{Domi10} 
  A.~Dom\'inguez,
  \textit{Capillary Forces between Colloidal Particles at Fluid Interfaces},
  in \textit{Structure and Functional Properties of Colloidal Systems},
  edited by R. Hidalgo-Alvarez
  (CRC Press, Boca Raton, FL, 2010), p.~31.

\bibitem{DaKr10a} 
  K.D.~Danov and P.~A.~Kralchevsky, 
  Adv.~Coll.~Interface Sci.~\textbf{154}, 91 (2010).

\bibitem{BLCS12} 
  L. Botto, E. P. Lewandowski, M. Cavallaro Jr., and K. J. Stebe,
  Soft Matter {\bf 8}, 9957 (2012).

\bibitem{CDR09}
  A. Campa, T. Dauxois, and S. Ruffo,
  Phys. Rep. \textbf{480}, 57 (2009).

\bibitem{Keller:1970}
  E.~Keller and L.~A.~Segel,
  J. Theor. Biol. \textbf{26}, 399 (1970).

\bibitem{ChSi08}
  P.-H.~Chavanis and C.~Sire,
  Physica A \textbf{387}, 4033 (2008).

\bibitem{Chavanis:2010}
  P.-H.~Chavanis,
  Physica A \textbf{390}, 1546 (2011).


\bibitem{Dominguez:2010}
  A.~Dom\'\i nguez, M.~Oettel, and S.~Dietrich, 
  Phys. Rev. E \textbf{82}, 011402 (2010).

\bibitem{Bleibel:L2011}
  J.~Bleibel, S.~Dietrich, A.~Dom\'\i nguez, and M.~Oettel, 
  Phys. Rev. Lett. \textbf{107}, 128302 (2011).

\bibitem{Bleibel:2011}
  J.~Bleibel, A.~Dom\'\i nguez, M.~Oettel, and S.~Dietrich 
  Eur. Phys. J. E \textbf{34}, 125 (2011).

\bibitem{ODD05}
  M. Oettel, A. Dom{\'\i}nguez, and S. Dietrich,
  Phys. Rev. E \textbf{71}, 051401 (2005).

\bibitem{Dominguez:2008}
  A.~Dom\'\i nguez, M.~Oettel, and S.~Dietrich, 
  J. Chem. Phys. \textbf{128}, 114904 (2008). 

\bibitem{Bleibel:L2013}
  J.~Bleibel, A.~Dominguez, F.~G\"unther, J.~Harting, M.~Oettel,
  arXiv:1305.3715 [cond-mat.soft].

\bibitem{Mar99} U.~M.~B.~Marconi and P.~Tarazona,
  J.~Chem.~Phys. {\bf 110}, 8032 (1999).

\bibitem{Chav11}
  P.-H.~Chavanis,
  Phys. Rev. E \textbf{84}, 031101 (2011).

\bibitem{ChSi11}
  P.-H.~Chavanis and C.~Sire,
  Phys. Rev. E \textbf{83}, 031131 (2011).

\bibitem{Gnedin:1998}
  N.~Y.~Gnedin and L.~Hui,
  Mon. Not. R. Astron. Soc. \textbf{296}, 44 (1998).

\bibitem{Buch95a}
  T.~Buchert, \textit{Lagrangian Perturbation Approach to the
    Formation of Large--Scale Structure}, in \textit{Proceedings of
    the International School of Physics ``Enrico Fermi''}, edited by
  S. Bonometto, J.R. Primack, A. Provenzale (IOS, Amsterdam,
  1997), \tr{vol.~132,} p.~543; \korr{see also \texttt{arXiv:astro-ph/9509005}}.

\bibitem{CoHi89}
  \tr{R. Courant and D. Hilbert, 
    \textit{Methods of Mathematical Physics, Vol.~II}
    (Wiley, \korr{New York}, 1989).}

\bibitem{AdBu99}
  S.~Adler and T.~Buchert,
  Astron. Astrophys. \textbf{343}, 317 (1999) 

\bibitem{Hunt62}
  C. Hunter,
  Astrophys.~J. \textbf{136}, 594 (1962).

\bibitem{ALP88}
  S.~J. Aarseth, D.~N.~C. Lin, and J.~C.~B. Papaloizou,
  Astrophys.~J. \textbf{324}, 288 (1988).

\bibitem{SaCo95}
  V. Sahni and P. Coles,
  Phys.~Rep. \textbf{262}, 1 (1995).


\end{thebibliography}
\end{document}